\documentclass[a4paper,fleqn,usenatbib]{mnras}

\usepackage[T1]{fontenc}
\usepackage{ae,aecompl}

\usepackage{graphicx}
\usepackage{amsmath}
\usepackage{amssymb}
\usepackage{epstopdf}
\usepackage{color}
\definecolor{linkblue}{rgb}{0,0,0.8}
\definecolor{linkgreen}{rgb}{0,0.5,0}
\hypersetup{pdfpagemode=UseNone, pdfstartview=FitH, linkcolor=linkblue, %
            citecolor=linkblue, urlcolor=linkblue}
\usepackage{widetext}
            
\def\beq{\begin{equation}}
\def\eeq{\end{equation}}
\def\nn{\nonumber}
\newcommand{\lp}{\left(}
\newcommand{\rp}{\right)}
\newcommand{\lb}{\left[}
\newcommand{\rb}{\right]}

\definecolor{purple}{rgb}{0.78,0.18,0.77}

\def\d{{\partial}}
\newcommand{\invMpc}{\,h\, {\rm Mpc}^{-1}\,}
\newcommand{\Omm}{\Omega_{\rm m}}

\newcommand\aowls{A_{\rm bar}}
\newcommand{\alphaowls}{\alpha_a^\text{(bar)}}
\newcommand{\Deltaowls}{\Delta_\mu^\text{(bar)}}
\newcommand{\fom}{\mathcal{F}}

\title[Cosmic Shear as a Probe of Galaxy Formation Physics]{Cosmic Shear as a Probe of Galaxy Formation Physics}

\author[Foreman, Becker, and Wechsler]{
Simon Foreman\thanks{E-mail: \href{sforeman@cita.utoronto.ca}{sforeman@cita.utoronto.ca}}$^{1,2,3,4}$,
Matthew R.\ Becker$^{1,2}$,
Risa H.\ Wechsler$^{1,2,4}$
\\
$^{1}$Kavli Institute for Particle Astrophysics and Cosmology, P.O.\ Box 2450, Stanford, CA 94305\\
$^{2}$SLAC National Accelerator Laboratory, 2575 Sand Hill Road, Menlo Park, CA 94025\\
$^{3}$Stanford Institute for Theoretical Physics, 382 Via Pueblo Mall, Stanford, CA 94305\\
$^{4}$Department of Physics, Stanford University, 382 Via Pueblo Mall, Stanford, CA 94305
}

\pubyear{2016}

\begin{document}
\label{firstpage}
\pagerange{\pageref{firstpage}--\pageref{lastpage}}
\maketitle

\begin{abstract}
We evaluate the potential for current and future cosmic shear measurements from large galaxy surveys to constrain the impact of baryonic physics on the matter power spectrum.  We do so using a model-independent parameterization that describes deviations of the matter power spectrum from the dark-matter-only case
as a set of principal components that are localized in wavenumber and redshift. We perform forecasts for a variety of current and future datasets, and find that at least $\sim$90\% of the constraining power of these datasets is contained in no more than nine principal components. The constraining power of different surveys can be quantified using a figure of merit defined relative to currently available surveys.
With this metric, we find that the final Dark Energy Survey dataset (DES Y5) and the Hyper Suprime Cam Survey will be roughly an order of magnitude more powerful than existing data in constraining baryonic effects.  Upcoming Stage IV surveys (LSST, Euclid, and WFIRST) will improve upon this by a further factor of a few.  We show that this conclusion is robust to marginalization over several key systematics.  
The ultimate power of cosmic shear to constrain galaxy formation is dependent on understanding systematics in the shear measurements at small (sub-arcminute) scales.  If these systematics can be sufficiently controlled, cosmic shear measurements from DES Y5 and other future surveys have the potential to provide a very clean probe of galaxy formation and to strongly constrain a wide range of  predictions from modern hydrodynamical simulations. 
\end{abstract}



\section{Introduction}

Weak gravitational lensing has tremendous potential to inform our knowledge of the universe, on distance scales ranging from galactic to cosmological. In particular, {\it cosmic shear}, the statistical measurement of the distortion of observed galaxy shapes due to lensing by the distribution of matter along the line of sight, can provide information about the background cosmological model, via its sensitivity to the growth of structure and the strength of gravitational clustering on various scales. The first successful measurements of cosmic shear were carried out using patches of the sky with area of order one square degree~\citep{2000MNRAS.318..625B,2000astro.ph..3338K,2000Natur.405..143W,2000A&A...358...30V}. More recent measurements from the Stripe 82 region of the Sloan Digital Sky Survey~\citep{2012ApJ...761...15L,2014MNRAS.440.1322H}, the Canada France Hawaii Telescope Lensing Survey (e.g.~\citealt{2013MNRAS.432.2433H}), the Dark Energy Survey Science Verification run~\citep{2015arXiv150705552T}, and the Kilo-Degree Survey~\citep{2015MNRAS.454.3500K} have used $\mathcal{O}(100)$ square degrees and are beginning to provide useful low-redshift cosmological constraints that are complementary to the high-redshift information obtained from the cosmic microwave background. Indeed, there currently exist discrepancies between the two sets of constraints that have yet to be resolved (see e.g.~\citealt{2015MNRAS.451.2877M} for further discussion). Future surveys like the Wide Field InfraRed Survey Telescope (WFIRST)\footnote{\href{http://wfirst.gsfc.nasa.gov}{http://wfirst.gsfc.nasa.gov}}, the Large Synoptic Survey Telescope (LSST)\footnote{\href{http://www.lsst.org}{http://www.lsst.org}}, and Euclid\footnote{\href{http://sci.esa.int/euclid}{http://sci.esa.int/euclid}} will expand the sky coverage to $\mathcal{O}(10000)$ square degrees, and measure the shapes of orders of magnitude more galaxies than previous observational programs.

The observed shapes of galaxies are altered by gravitational lensing by only about 1\%, and it is the statistical correlations among these shapes that are the primary end product of cosmic shear observations. There are many possible sources of systematic errors that could contaminate the underlying cosmological signal, and each must be characterized in detail in order to obtain robust results. These include instrumental distortions of observed galaxy shapes caused by the telescope's point-spread function; biases in the methods used to measure shapes from raw images; uncertainties in the photometric redshifts obtained for each galaxy; ``intrinsic alignments" of galaxies with nearby structures, independently of the alignments caused by lensing by the line-of-sight distribution of matter; and uncertainties in the theoretical modeling of the distribution of matter on small scales.

This paper will focus on the last point, in particular on the effects of baryonic physics on the underlying two-point statistics of the distribution of matter. It is well-known~\citep{2004APh....22..211W,2004ApJ...616L..75Z,2006ApJ...640L.119J,2008ApJ...672...19R} that these effects can strongly influence cosmological constraints derived from cosmic shear measurements if not properly accounted for. For this reason, it is common in cosmological analyses to simply exclude scales where these effects are important. Alternatively, a variety of schemes have been developed to allow inclusion of these scales by mitigating the effects of baryons on the desired information. These includes parametrizations of baryonic effects in the context of the halo model~\citep{2008PhRvD..77d3507Z,2011MNRAS.417.2020S,2013MNRAS.434..148S,2013PhRvD..87d3509Z,2014arXiv1410.6826M,2014MNRAS.445.3382M,2015MNRAS.454.1958M}, perturbation theory~\citep{2015JCAP...05..019L}, principal component analysis~\citep{2015MNRAS.454.2451E,2016MNRAS.459..971K}, or empirical fitting functions for the net effect~\citep{2015MNRAS.450.1212H} or individual baryonic components~\citep{2015JCAP...12..049S}, along with other techniques not based on direct parametrizations~\citep{2005PhRvD..72d3002H,2011MNRAS.416.1717K,2015JCAP...05..023B}.

However, as in~\citet{2015MNRAS.450.1212H}, one can also turn the question around, and treat these baryonic effects as {\it signal} to be constrained instead of nuisance to be removed or marginalized over. The advantage of using cosmic shear for these constraints is that it provides a relatively clean probe of the underlying matter distribution, free of the need to classify individual objects and explicitly relate them to dark matter halos or properties of their environments. Similarly, the effect of baryons on the total matter distribution can also be obtained relatively easily from a numerical simulation, by simply measuring the $N$-point matter statistics from snapshots of the simulation output. In recent years, such simulations have reached sufficient levels of detail that it is worthwhile to compare their predictions to observations of the real universe, and cosmic shear can act as nice complement to other observations that have already been exploited for this purpose (e.g.~\citealt{2010MNRAS.402.1536S,2015MNRAS.452.3529V}).

Rather than performing a separate analysis to constrain each implementation of baryonic physics one is interested in, it is possible to obtain generic, model-independent constraints via a principal component analysis of the deviations of the observed matter power spectrum from a fiducial ``dark matter-only" case. One can construct principal components of these deviations (in the form of linear combinations of deviations from the dark matter-only power spectrum at discrete points in the wavenumber-redshift plane) from a Fisher matrix corresponding to a specific survey: inverting and then diagonalizing this matrix not only gives the principal components corresponding to that survey, but also provides an automatic ranking of these principal components in terms of the expected constraints on their amplitudes. (For applications of this procedure to other problems in cosmology, see~\citealt{2003PhRvL..90c1301H,2009arXiv0901.0721A,2012PhRvD..86l3504S}.) One can then simply constrain these amplitudes in a likelihood analysis, and thereafter straightforwardly relate these constraints to any baryonic model of interest.

One may contrast this approach to that of~\citet{2015MNRAS.454.2451E}, which also applies principal component analysis to the issue of baryonic effects. The approach in that work is to find the principal components of the variations in the matter power spectrum between a range of specific baryonic models, and then remove these principal components from both the data and theory vectors in a likelihood analysis, in order to minimize the amount that cosmological constraints are contaminated by these effects. Our approach is rather to find the principal components of the matter power spectrum that will be best constrained by a cosmic shear measurements from a given survey, and then relate constraints on these principal components to constraints on a set of baryonic models as a second independent step.

In this paper, we carry out forecasts to assess the performance of this approach when it is applied to current and upcoming weak lensing surveys. Our forecasts marginalize over a variety of systematic effects relating both to calibration of the shear measurements and uncertainties in the theoretical modeling. For all surveys we consider, we find that 90\% of the constraining power contained in the shear correlation function is captured by at most nine principal components of the matter power spectrum. We use this fact to define a  figure of merit:  the reciprocal of the geometric mean on the expected one-sigma constraints of the amplitudes of the nine principal components. Comparing this figure of merit for different surveys, we find that the constraints on baryonic effects from Stage III surveys like the Dark Energy Survey or Hyper Suprime Cam Survey\footnote{\href{http://www.naoj.org/Projects/HSC/}{http://www.naoj.org/Projects/HSC/}} can improve upon those from currently available datasets by roughly an order of magnitude, while Stage IV surveys will only improve upon this by a factor of a few, with the constraining power being driven mainly by the number of galaxy shapes that can be measured by each survey.

We then apply this method to a representative set of models for baryonic effects on the matter power spectrum; specifically, we use nine models from the OverWhelmingly Large Simulations (OWLS;~\citealt{2010MNRAS.402.1536S,2011MNRAS.415.3649V}) suite. We find that the constraining power of future surveys is strongly dependent on the minimum angular scale at which measurements of the shear correlation functions can be used. In particular, measurements at scales of less than one arcminute would allow Stage III surveys to rule out the majority of the OWLS models at more than five-sigma confidence (or alternatively rule out the dark matter-only power spectrum if one of these models were correct). Furthermore, these measurements would also likely allow for the identification of a single ``best-fit" model among those that currently exist. 

While we marginalize over a wide range of systematics, our conclusions rely on the existence of shear catalogs with well-understood properties (in terms of, for example, selection functions) at the appropriate scales. Provided that such an understanding can be attained (particularly for $\theta\lesssim 1'$), however, our results demonstrate that cosmic shear is capable of providing stringent tests of the results of current and future hydrodynamical simulations, opening an additional window on the physics of galaxy formation and evolution (or at least, its implementation in simulations).

The paper is organized as follows. In Sec.~\ref{sec:pca}, we describe the principal component method we use for parameterizing the matter power spectrum, while in Sec.~\ref{sec:mockdataandbaryons} we describe our assumptions about the surveys we provide forecasts for, the mock cosmic shear data and covariances we associate with these surveys, and the set of baryonic models we will use for our tests. In Sec.~\ref{sec:valopt}, we verify that most of the constraining power of each survey with respect to the matter power spectrum is contained in no more than the first nine principal components, and describe how we relate constraints on these principal components to a given baryonic model. In Sec.~\ref{sec:forecasts}, we compare the performance of various surveys via an appropriately-defined figure of merit, and also quantify this performance in terms of constraints on the baryonic models in our test set. We additionally examine the impact of varying our priors on the most important theoretical systematics. Finally, we provide additional comments and conclude in Sec.~\ref{sec:conclusions}.

\section{Method of Principal Components}
\label{sec:pca}

\begin{figure*}
\includegraphics[width=0.99\textwidth]{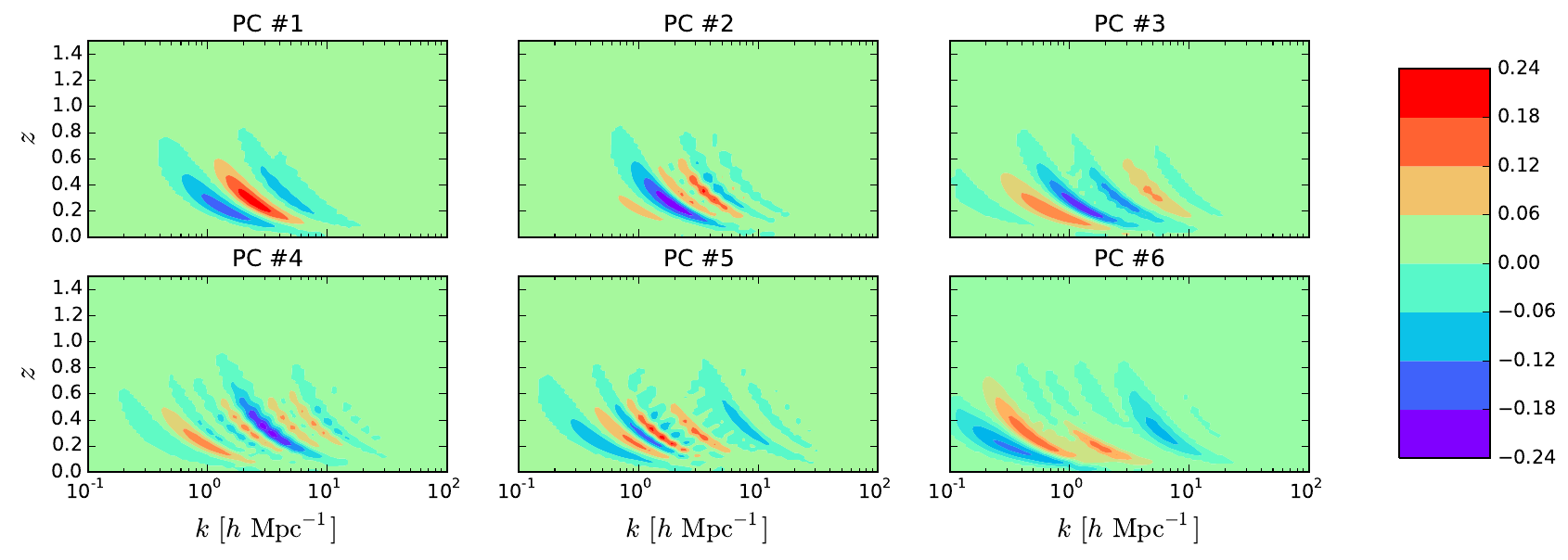}
\caption{\label{fig:example_modes}
A visualization of the first six principal components of $\Delta(k,z) \equiv P(k,z)/P_\text{DM-only}(k,z)-1$ that will be best constrained by measurements of the cosmic shear correlation functions $\xi_\pm$ from DES Y5.
(See Sec.~\ref{sec:mockdataandbaryons} for our assumptions about the details of these measurements.) We use ``principal component" to refer to a linear combination of values of~$\Delta$ defined on a grid in wavenumber and redshift, and the color scheme indicates how different regions in this grid are weighted in each linear combination (see Sec.~\ref{sec:pca} for details).
The shapes and localization of these principal components correspond to curves in the $(k,z)$ plane that contribute to multipoles at different $\ell$ values.
}    
\end{figure*}

We begin by seeking a generalized parametrization of the matter power spectrum that allows for deviations from the DM-only case that are localized in wavenumber $k$ and redshift $z$. Such a parametrization will allow us to account for the fact that the strongest constraints on $P(k,z)$ from cosmic shear will be localized where the lensing kernels peak, and also for the fact that models for baryonic effects will have characteristic scale- and redshift-dependence that will vary from model to model.

One such parametrization simply uses the fractional deviation $\Delta$ between the full power spectrum and the DM-only one at discrete values of $k$ and $z$, taking the value of that deviation at each sample as a free parameter and smoothly interpolating between these samples at other values of $k$ and~$z$:
\beq
P(k_i,z_j) = P_\text{DM-only}(k_i,z_j) \lb 1+\Delta_{ij} \rb\ .
\eeq
A version of this parametrization that assumes that $\Delta$ is $z$-independent was used in~\citet{2005APh....23..369H} and~\citet{2012JCAP...04..034H}. (For other works that do not rely on a $z$-independent parametrization, see~\citealt{2009ApJ...695..652B}, \citealt{2015JCAP...05..023B}, and particularly~\citealt{2012A&A...543A...2S}, which presents estimators for the $\{\Delta_{ij}\}$ from a set of two-point function measurements.) For our purposes, it is important to relax this assumption, but this requires the introduction of at least several tens of new parameters $\{\Delta_{ij}\}$ into any analysis that would make use of this parametrization. Following the discussion above, some of these parameters will be much better constrained by a particular dataset than others. In addition, the nature of cosmic shear observables as projections of the matter power spectrum will lead to strong degeneracies among these parameters, since there are many ways to perturb the $\{\Delta_{ij}\}$ in a correlated way without significantly affecting the final signal (although useful constraints can be obtained even in the presence of such degeneracies---see~\citealt{2012A&A...543A...2S}).

Therefore, we further seek a convenient basis for these parameters that eliminates degeneracies as much as possible, and also identifies which basis elements are most relevant for a given set of observations. In fact, such a basis is straightforward to construct: given a covariance matrix for the $\{\Delta_{ij}\}$, its eigenvectors would specify a set of linear combinations of the $\{\Delta_{ij}\}$ that are maximally uncorrelated with one another, while the corresponding eigenvalues would indicate the expected variance on a measurement of each linear combination. Sorting these ``principal components" (PCs) by eigenvalue, we can then hope to identify a minimal subset of them that will suffice for a particular analysis (in our case, providing sufficient constraints on a range of baryonic models). Such a procedure has previously been applied, for example, to the reconstruction of a time-dependent equation of state for dark energy~\citep{2003PhRvL..90c1301H}, and of the primordial spectrum of curvature perturbations from inflation~\citep{2005PhRvD..72b3510K,2010PhRvD..82d3513D}, and~\citet{2012A&A...543A...2S} also suggested that an application to the matter power spectrum might be useful.

For a given survey, the covariance matrix of the $\{\Delta_{ij}\}$ (along with other parameters related to the background cosmology and systematics) can be approximated by the inverse of the appropriate Fisher matrix. Recall that, in general, for a given vector of observed data points $\boldsymbol{d}$ with associated covariance~$\mathbf{Cov}$, along with a given model with parameters~$\{p_a\}$, the Fisher matrix is given by\footnote{The Fisher matrix will also contain terms that include the derivative of $\mathbf{Cov}$ with respect to the model parameters, but we will ignore those terms because they will generally be subdominant for a large-area survey (e.g.~\citealt{2009A&A...502..721E}).}
\beq
F_{ab} = \frac{\d\boldsymbol{d}^{\rm T}}{\d p_a} \mathbf{Cov}^{-1} \frac{\d\boldsymbol{d}}{\d p_b}
	+ \frac{\delta^{ab}}{\sigma_a^2}\ ,
\eeq
where the last term incorporates optional Gaussian priors of width $\sigma_a$ on each parameter~$p_a$. In our case, $\boldsymbol{d}$ will be a vector of binned shear correlation function measurements, $\mathbf{Cov}$ will be their covariance (estimated either from mocks or analytical modeling), and the model parameters~$\{p_a\}$ are the $\{\Delta_{ij}\}$, along with any other cosmological or systematics parameters that would be varied in a likelihood analysis. Upon inverting $F_{ab}$ and then diagonalizing the submatrix of $\mathbf{F}^{-1}$ corresponding to the $\{\Delta_{ij}\}$, we obtain a new set of parameters~$\{\alpha_a\}$, defined by
\beq
\alpha_a = \sum_\mu \beta_{a\mu} \Delta_\mu\ ,
\label{eq:alphaa_def}
\eeq
where the $(i,j)$ indices on $\Delta$ have been compressed into a single index $\mu$ for brevity. Thus, each~$\alpha_a$ is the value of a certain linear combination of $\Delta_\mu$ values, with each linear combination defined by the coefficients~$\{\beta_{a\mu}\}$. For example, if $\beta_{1\mu}$ were equal to $(\delta_{\mu1}+\delta_{\mu2})/\sqrt{2}$, then~$\alpha_1$ would be the value of $(\Delta_1+\Delta_2)/\sqrt{2}$ inferred from the data. We use conventions where the $\{\boldsymbol{\beta}_{a}\}$ are orthonormal vectors, i.e.~$\sum_\mu \beta_{a\mu} \beta_{b\mu} = \delta_{ab}$. We will denote the eigenvalues (forecast variances of each $\alpha_a$) by $\sigma_a^2$.

\begin{figure}
\includegraphics[width=\columnwidth]{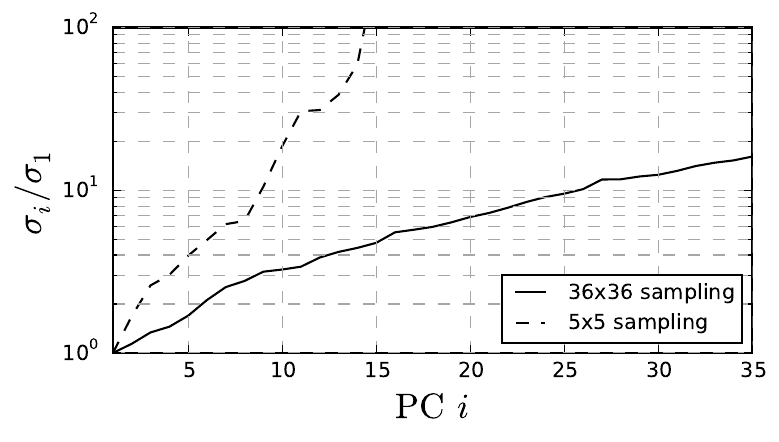}
\caption{\label{fig:example_sigmas}
Expected one-sigma uncertainties $\sigma_i$ on the amplitudes $\alpha_i$ of the most relevant PCs for DES Y5, normalized to the uncertainty on the best-constrained PC. The solid line corresponds to PCs constructed from a fine sampling of the $(k,z)$ plane (a $36\times36$ grid in $\log(k/[\invMpc])$ and $z$), while the dashed line corresponds to a coarser sampling (a $5\times5$ grid). The expected uncertainties increase rapidly with mode number, implying that a finite number of the lowest modes likely contain the bulk of a survey's information about the matter power spectrum. Furthermore, this increase is most rapid when the $(k,z)$ plane is sampled more coarsely, since each sample then acts as a wider ``bin" in the~$(k,z)$ plane and thus encapsulates more information. Forecasts for other surveys exhibit the same behavior.
}    
\end{figure}

In Fig.~\ref{fig:example_modes}, we visualize the results of this procedure for the six PCs whose amplitudes will be best constrained by (i.e.\ have the lowest variance when constrained using) measurements of the cosmic shear two-point function from DES Y5, where the assumed properties and treatment of systematics for DES Y5 will be described in Secs.~\ref{sec:mockdata}-\ref{sec:systematics}. Specifically, for each PC indexed by $a=1\dots 6$, we plot the values of the $\{\beta_{a\mu}\}$ mapped to the appropriate location in the $(k,z)$ plane, smoothly interpolating between grid points and representing the numerical $\beta_{a\mu}$ values using the color scheme indicated to the right of the figure.
The PCs in this figure were constructed from a set of $\{\Delta_{ij}\}$ defined in a $36\times36$ grid in $\log(k/[\invMpc])$ and $z$, bounded by the edges of each panel in Fig.~\ref{fig:example_modes}. The lower bound for $k$, $0.1\invMpc$, has specifically been chosen to exclude linear scales, while the other boundaries have been chosen to encompass the range of wavenumbers and redshifts where DES is expected to be most informative (assuming measurements of $\xi_\pm$ to $\theta_{\rm min}=2'$). Similar considerations apply to our forecasts for other surveys.

The appearance of these modes in Fig.~\ref{fig:example_modes} reflects the sensitivity of the data to projections of $P(k,z)$ that contribute to different angular scales: specifically, a given multipole~$C_\ell$ will probe the matter power spectrum along the curve $k=\ell/\chi(z)$, where $\chi(z)$ is the comoving distance to redshift $z$ (see Eq.~\eqref{eq:cl_conv} in Sec.~\ref{sec:mockdata}). Measurements of $\xi_\pm$ in a given angular bin will be sensitive to a range of multipoles; it is also possible that sets of angular bins exist that are impacted equally by fractional variations in the power spectrum at a given level. Fig.~\ref{fig:example_modes}  demonstrates that these angular bins (and their corresponding ranges of multipoles) are identified and combined together automatically by the PCA procedure.

\begin{figure}
\begin{centering}
\includegraphics[width=\columnwidth]{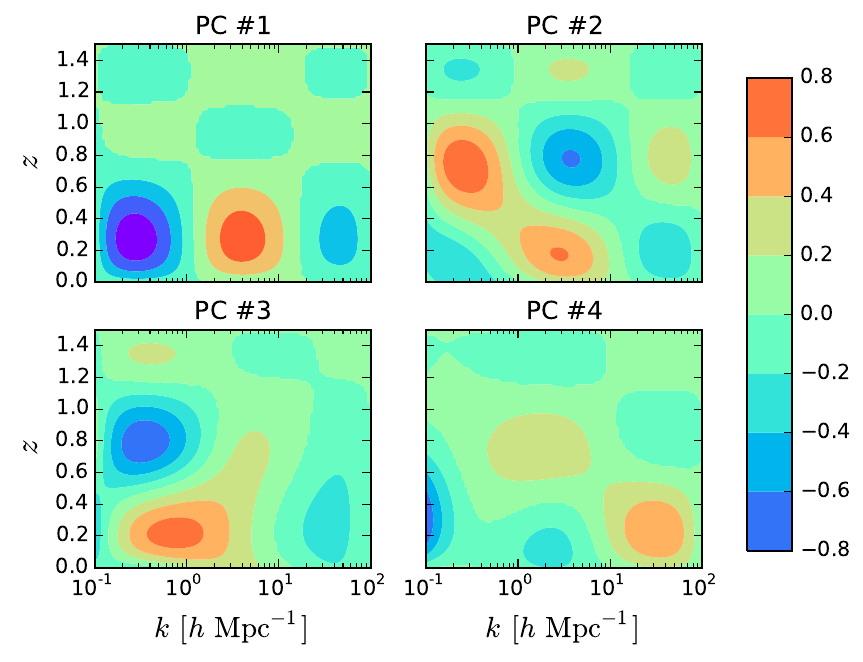}
\caption{\label{fig:example_modes_coarse}
As Fig.~\ref{fig:example_modes}, but displaying PCs constructed from a set of $\{\Delta_{ij}\}$ that sample the $(k,z)$ plane a factor of $\sim$5 more coarsely in both $k$ and $z$. With this coarser sampling, measurements at a greater range of angular scales are important for each PC, with the result that these four PCs contain a much larger fraction of the information to be found in DES than the first four PCs shown in Fig.~\ref{fig:example_modes}. For this reason, we will use a similarly coarse sampling of the $(k,z)$ plane in the remainder of this work.
}    
\end{centering}
\end{figure}

In Fig.~\ref{fig:example_sigmas}, we show the forecasted $1\sigma$ uncertainties on the amplitudes of the most important PCs for DES Y5, normalized to the uncertainty on the PC expected to be best constrained. (These uncertainties are simply the square roots of the eigenvalues produced by the diagonalization procedure described above.)
The solid line in this figure uses PCs obtained from the same $36\times36$ grid in $\log(k/[\invMpc])$ and $z$ for $\Delta_{ij}$ that was used to produce Fig.~\ref{fig:example_modes}, while the dashed line arises from a much coarser $5\times5$ grid in $\log(k/[\invMpc])$ and $z$. (The four best-constrained PCs corresponding to this coarser sampling are shown in Fig.~\ref{fig:example_modes_coarse}.) 

The steep dropoff in constraining power (generically also seen in forecasts for other surveys, not shown in Fig.~\ref{fig:example_sigmas}) indicates that a finite number of PCs will likely contain the majority of the information about the matter power spectrum that we could expect to extract from a given set of shear measurements. The difference between the dashed and solid lines further implies that a coarser sampling of the $(k,z)$ plane is more efficient in compressing the constraining power into a small set of PCs, since a coarser sampling effectively bins more of the signal in the $(k,z)$ plane into each grid point. Henceforth, in this work we will use a $5\times5$ grid to produce our results, since this is the smallest number of points that will allow us to use cubic splines to interpolate along each axis. Coarser grids paired with different interpolation methods may also be useful, but we will not pursue them in this work.

It is worth mentioning that in the original basis of~$\{\Delta_{ij}\}$ variables, some of these variables may be so poorly constrained that their corresponding elements in the Fisher matrix will cause the matrix to become ill-conditioned, impeding its inversion or diagonalization. This can be ameliorated by incorporating a large Gaussian prior for each variable, which will not affect the results for the most important modes. For our forecasts in this work, a prior of $\sigma(\Delta_{ij})=10^4$ was sufficient to eliminate these numerical issues.

\begin{table*}
	\centering
	\caption{Assumed properties of surveys considered in this work}
	\label{tab:survey_info}
	\begin{tabular}{lccccc}
		\hline
		 & Sky coverage & Source density, & Number of & Source distribution, \\
		 & (deg$^2$) & $\bar{n}$ (gals/arcmin$^2$) & redshift bins & $n(z)$  \\
		\hline
		DES SV	& 139	& 5.7		& 3 
			& From~\citep{2015arXiv150705598B}		\\
		DES Y5	& 5000 	& 8		& 5 
			& Eq.~\eqref{eq:nz}: $z_{\rm max}=2.0$, $\alpha=1.75$, $z_0=0.25$, $\beta=1.0$		\\
		HSC		& 1400 	& 20		& 10 
			& Eq.~\eqref{eq:nz}: $z_{\rm max}=3.0$, $\alpha=2.0$, $z_0=0.35$, $\beta=1.0$		\\
		LSST	& 18000  	& 26		& 10 
			& Eq.~\eqref{eq:nz}: $z_{\rm max}=3.5$, $\alpha=1.25$, $z_0=0.50$, $\beta=1.0$		\\
		Euclid	& 15000 	& 30		& 10 
			& Eq.~\eqref{eq:nz}: $z_{\rm max}=2.5$, $\alpha=1.25$, $z_0=0.35$, $\beta=1.0$		\\
		WFIRST	& 2200 	& 45		& 10 
			& Eq.~\eqref{eq:nz}: $z_{\rm max}=4.0$, $\alpha=1.25$, $z_0=0.60$, $\beta=1.0$		\\
		\hline
	\end{tabular}
\end{table*}

\section{Mock Data and Baryonic Models}
\label{sec:mockdataandbaryons}

We will demonstrate the performance of the method described in Sec.~\ref{sec:pca} by carrying out forecasts in which it is applied to a number of mock data sets corresponding to current and upcoming weak lensing surveys, and comparing to what would be obtained from attempting to directly constrain a set of representative models for baryonic effects on the matter power spectrum. In this section, we describe how these mock data sets are constructed, our treatment of systematic effects that could be important for cosmic shear measurements on the relevant scales, and the baryonic models we use for our test set.

\subsection{Mock data}
\label{sec:mockdata}

For our forecasts, we consider a variety of upcoming surveys (along with the already-completed Dark Energy Survey Science Verification run, DES SV, which will serve as an indicator of the constraining power of currently available data). Table~\ref{tab:survey_info} displays the properties we will assume in this work. These properties are based mainly on~\citet{2016MNRAS.456..207K}, with following exceptions: for DES Y5, we have use a slightly more conservative value for the density of sources; for HSC, we have used the sky coverage and a more conservative value of the source density from~\citet{2015ApJ...806..186O}; and for Euclid, we have used the density of sources specified in~\citet{2011arXiv1110.3193L}. For all surveys, we will assume an intrinsic shape noise of $\sigma_\epsilon=0.27$ per component. Where appropriate, throughout the paper we will comment on the sensitivity of our forecasts to these assumptions.

For the data themselves, we will use the real-space two-point shear correlation functions $\xi_\pm(\theta)$, given by 
\beq
\xi_{\pm}^{ij}(\theta) = \frac{1}{2\pi} \int d\ell \, \ell \, J_{0/4}(\ell\theta) \, C_\ell^{ij}\ ,
\eeq
where $J_{0/4}$ is the Bessel function of order 0 or 4, and $i,j$ denote the two tomographic redshift bins involved in the correlation. The multipoles of the convergence field, $C_\ell^{ij}$, are obtained from the matter power spectrum via
\beq \label{eq:cl_conv}
C_\ell^{ij} = \frac{9H_0^4 \Omm^2}{4c^4} \int_0^{\chi_{\rm h}} d\chi
	\frac{g^i(\chi) g^j(\chi)}{a^2(\chi)} \,P\!\lp \frac{\ell}{f_K(\chi)}, z(\chi) \rp\ ,
\eeq
where $\chi_{\rm h}$ is the comoving distance to the horizon and $f_K(\chi)$ is the comoving angular diameter distance (equal to $\chi$ in a flat universe, which we will assume in this work). The lens efficiency $g_i(\chi)$ is given by
\beq \label{eq:lens_eff}
g^i(\chi) = \int_\chi^{\chi_{\rm h}} d\chi' n_i(\chi') \frac{f_K(\chi'-\chi)}{f_K(\chi')}\ .
\eeq
Our base forecasts will be performed using 6 measurements of each of $\xi_+$ and $\xi_-$ per redshift bin pair, log-spaced between $\theta_{\rm min}=2'$ and $\theta_{\rm max}=300'$, although we will also investigate the consequences of other choices of $\theta_{\rm min}$.

For the distribution of sources $n_i(z)$ in each redshift bin, we will use the exact distributions from DES SV~\citep{2015arXiv150705598B} for the corresponding forecast. For DES Y5 and the other surveys, we employ the following parametrization with parameter values given in Table~\ref{tab:survey_info}, dividing the total $n(z)$ into bins with equal numbers of sources:
\beq \label{eq:nz}
n(z) \propto \Theta(z_{\rm max}-z) \, z^\alpha \exp\lb -\lp \frac{z}{z_0} \rp^\beta \rb\ .
\eeq
We base the source distribution for HSC  on~\citet{2011PhRvD..83b3008O}, while we follow~\citet{2016MNRAS.456..207K} for the other surveys. We make a conservative choice of 5 redshift bins for DES Y5, while we will assume that the other surveys will be capable of collecting sufficient statistics in 10 redshift bins.

For the covariance of the mock correlation function measurements, we will restrict ourselves to the Gaussian approximation, evaluated as in~\citet{2008A&A...477...43J}. We refer the reader to that paper for the derivation, and merely quote the final expression we use, which accounts for the difference in our conventions for $\sigma_\epsilon$:
\begin{align} \nn
&\text{Cov}_\text{G} \! \lb \xi_\pm^{ij}(\theta_1), \xi_\pm^{kl}(\theta_2)  \rb \\ \nn
&\quad = \frac{1}{8\pi^2 f_{\rm sky}} 
	\int d\ell \, \ell\, J_{0/4}(\ell\theta_1) J_{0/4}(\ell\theta_2) \\ \nn
&\qquad\qquad\qquad\times
	\lb \tilde{C}_{\ell_{1}}^{ik} \tilde{C}_{\ell_{1}}^{jl} + \tilde{C}_{\ell_{1}}^{il} \tilde{C}_{\ell_{1}}^{jk} 
	- (\delta^{ik} \delta^{jl} + \delta^{il}\delta^{jk}) \frac{\sigma_\epsilon^4}{n^2} \rb \\
&\qquad + (\delta^{ik} \delta^{jl} + \delta^{il}\delta^{jk}) \delta_{\theta_1 \theta_2}
	\delta_{+-}
	\frac{\sigma_\epsilon^4}{8\pi^2 f_{\rm sky} \theta_1 \Delta\theta_1 n^2 }\ ,
\label{eq:xipm_cov}
\end{align}
\noindent where $\delta_{+-}=1$ for $\text{Cov}_\text{G} \! \lb \xi_+, \xi_+ \rb$ or $\text{Cov}_\text{G} \! \lb \xi_-, \xi_- \rb$ and 0 otherwise, and $n$ is the angular density of source galaxies in each redshift bin (in our forecasts, $n$ is the same for each bin in a given survey). In Eq.~\eqref{eq:xipm_cov}, we have defined
\beq
\label{eq:clplussn}
\tilde{C}_{\ell}^{ij} = C_{\ell}^{ij} + \delta^{ij} \frac{\sigma_\epsilon^2}{n}\ .
\eeq

For a subset of our forecasts, we have also implemented the leading non-Gaussian terms in the covariance (the trispectrum of the convergence, along with a halo sample variance term that accounts for the influence of modes beyond the survey window), using the halo model treatment described in~\citet{2015MNRAS.454.2451E}, to which we refer the reader for details. We find that the inclusion of these terms has negligible impact on our results, justifying our use of the Gaussian approximation for the remaining forecasts. 

For our fiducial cosmology, we use the following best-fit parameters from~\citet{2015arXiv150201589P}: $\Omm=0.316$, $\Omega_{\rm b}=0.0491$, $h_0=0.673$, $\sigma_8=0.83$, $n_{\rm s}=0.965$, $w=-1.0$, and $\tau=0.078$. In our forecasts, we allow these parameters to vary within the one-sigma priors determined from Planck. Previous work~\citep{2014PhRvD..90f3516N,2015MNRAS.450.1212H} has shown that it is important to consider the effect of massive neutrinos jointly with other effects on the small-scale matter power spectrum, and therefore we also marginalize over the sum of neutrino masses in our forecasts, assuming a fiducial value of $\Sigma m_\nu = 0.06$eV (for simplicity, we only assume a single massive neutrino species).

For each survey, we compute a mock data vector and mock covariance matrix with the properties and fiducial cosmology described above. We then use these to compute a Fisher matrix, where the set of varied parameters consists of the cosmological parameters in the previous paragraph, neutrino mass, the grid of 25 $P(k,z)$ values described in Sec.~\ref{sec:pca}, and the systematics parameters listed in the next section. The maximum redshift for the power spectrum grid is set roughly to the maximum redshift for each survey, while the maximum wavenumber is set based on the value $\theta_{\rm min}$.\footnote{Specifically, we use $k_{\rm max}=\{ 50, 100, 400 \} \invMpc$ for $\theta_{\rm min} = \{4',2',0.5'\}$. These $k_{\rm max}$ values we chosen to obtain convergence of the final forecast results with respect to small changes in $k_{\rm max}$. Therefore, the results are not strongly sensitive to the values of the power spectrum precisely at $k_{\rm max}$, but are only sensitive up to a wavenumber around a factor of a few smaller.} We then invert the Fisher matrix and diagonalize the submatrix corresponding to the $P(k,z)$ samples; the eigenvectors (PCs) and eigenvalues $\{\sigma_a^2\}$ are then used to obtain the results in the remainder of the paper.

We use CAMB~\citep{camb} to calculate the linear matter spectrum, and Halofit~\citep{halofit,halofit-update} as an approximation for the nonlinear corrections from gravitational evolution. Furthermore, we use the modified Halofit prescription from~\citet{2012MNRAS.420.2551B} to account for the effect of massive neutrinos. (In Sec.~\ref{sec:systematics}, we describe how we account for possible uncertainties in the modeling of the nonlinear power spectrum.) Most numerical computations are performed using CosmoSIS~\citep{2015A&C....12...45Z}, with the exception of the mock covariances, for which we use CosmoLike~\citep{2016arXiv160105779K}.

\subsection{Treatment of systematic errors}
\label{sec:systematics}

We have attempted to include a selection of possible systematic errors that will likely be relevant for cosmic shear measurements on the scales we are interested in.
Our implementation is as follows:
\begin{enumerate}
\item {\bf Shear calibration errors}: 
Following the treatment of the DES Science Verification Data in~\citet{2015arXiv150705552T}, we assign a single free multiplicative parameter $m_i$ to each redshift bin, so that $C_\ell^{ij} \to (1+m_i)(1+m_j) C_\ell^{ij}$, and marginalize over each of these parameters, using a Gaussian prior of width 0.05 centered at zero. This choice is extensively discussed and motivated in~\citet{2015arXiv150705603J} in the context of DES SV, but the general features of that discussion are likely to apply to shear calibration issues in other surveys as well.
\item {\bf Photometric redshift uncertainties}: 
For each redshift bin, we allow for a uniform translation $\delta z_i$ of the entire source distribution within that bin, such that $n_i(z) \to n_i(z-\delta z_i)$. We marginalize over each $\delta z_i$ with a prior of 0.05, which was found to reflect the uncertainty in the redshift distributions from DES SV~\citep{2015arXiv150705909B}; while this value is much larger than the uncertainty goal for photo-$z$'s from future (e.g.\ Stage III) observations, to be conservative, we use it for all surveys we consider.
\item {\bf Intrinsic alignments}: 
We assume the non-linear tidal alignment model of~\citet{2007NJPh....9..444B}, with a single free overall amplitude $A_{\rm IA}$, assigned a wide prior of $\sigma(A_{\rm IA})=5$ and marginalized over in our analysis. 
While the choice of intrinsic alignment model could contribute as an additional source of uncertainty,~\citet{maccrann-inprep} find that the use of a more general model does little to degrade the constraints on baryonic physics (in that work parametrized through the halo model of~\citealt{2015MNRAS.454.1958M}), and we anticipate that this conclusion will also apply to the method we employ here.
\item {\bf Uncertainty in power spectrum modeling}: 
We treat this uncertainty by allowing for two constant fractional shifts in the value of the ``DM+neutrinos" power spectrum, above and below some transition wavenumber $k_{\rm tr}$, with exponentials smoothly connecting the two regimes:
\begin{align} \nn
&P(k,z) \to P(k,z) \lb 1 + S_P^{\rm (low)} \exp({-k^2/k_{\rm tr}^2}) \right.  \\
&\qquad\qquad\qquad\qquad\quad \left. +\,\, S_P^{\rm (high)} \lp 1-\exp[{-k^2/k_{\rm tr}^2}] \rp \rb\ .
\end{align}
We fix $k_{\rm tr}=0.5\invMpc$, and marginalize over the two amplitudes $S_P^{\rm (low)}$ and $S_P^{\rm (high)}$, with Gaussian priors of 0.02 and 0.05 respectively (both centered at zero). These priors are roughly based on conservative estimates of the accuracy of currently-available fitting functions like Halofit or the CosmicEmu emulator~\citep{2014ApJ...780..111H}. In the future, the combination of improvements in perturbation theory~\citep[e.g.][]{2015arXiv150705326F} at large scales and more ambitious emulation programs or approaches to nonlinear modeling at smaller scales will likely decrease these errorbars dramatically. We explore the effect of varying the prior on $S_P^{\rm (high)}$ in Sec.~\ref{sec:sys-priors} (our results are insensitive to the prior on $S_P^{\rm (low)}$).
\item {\bf Other small-scale shear systematics}:
There exist several other systematic effects that could become relevant at high multipoles, or equivalently small ($\sim$arcminute) angular scales. Examples of these include higher-order terms in the expansion of the ``reduced shear" $\gamma/(1-\kappa)$, which is what is actually observed (rather than the pure shear $\gamma$ itself)~\citep{2009ApJ...696..775S,2010A&A...523A..28K}; {\it lensing bias}, the effect of lensing on observed properties of galaxies, which can lead to biased selection in regions of large magnification~\citep{2009ApJ...702..593S,2010A&A...523A..28K}; and environment-dependent selection biases, arising for example from excluding blended sources (which typically occur in regions of high convergence) from a shear catalogue~\citep{2011A&A...528A..51H,maccrann-thesis,maccrann-inprep}.

Fortunately, a simple power-law model, $\Delta C_\ell/C_\ell \propto \ell^p$ serves to mimic the impact of most of these effects quite closely, with  $p \approx 0.5$ providing a good match to examples of explicit calculations found in the literature~\citep{2009ApJ...696..775S,2009ApJ...702..593S}. Therefore, we roll these effects into the following phenomenological model, letting $S_{\rm shear}$ be the overall amplitude of these effects at $\ell=10^4$:
\beq
C_\ell^{ij} \to C_\ell^{ij} \lb 1 + S_{\rm shear} \lp \frac{\ell}{10^4}\rp^{0.5} \rb\ .
\eeq
Such a model was also used in~\citet{2006MNRAS.366..101H}, who pointed out that other additive systematics such as anisotropies in a telescope's point-spread function will likely have a similar impact on the angular power spectrum.

We marginalize over the amplitude $S_{\rm shear}$ with a prior of 0.05, which is roughly the level at which these effects are expected to contaminate near-term shear measurements~\citep{maccrann-thesis,maccrann-inprep}. We investigate how our results depend on the prior on $S_{\rm shear}$ in Sec.~\ref{sec:sys-priors}.
\end{enumerate}

In summary, we marginalize over $(2n_{\rm bins}+4)$ systematics parameters, along with the sum of neutrino masses.

\subsection{Models for baryonic effects}
\label{sec:baryon_models}

\begin{figure}
\includegraphics[width=\columnwidth]{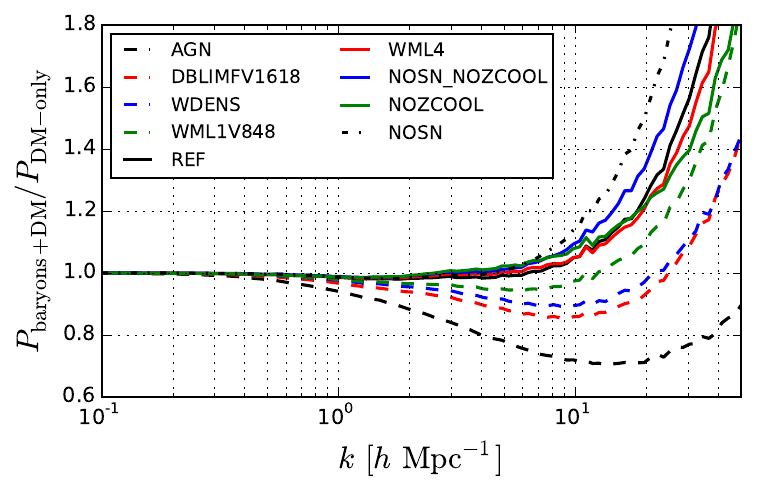}
\caption{\label{fig:baryons_pkratio}
Ratio of total (baryons+DM) power spectra over DM-only power spectrum  at $z=0$ for each baryonic model described in Sec.~\ref{sec:baryon_models}.
}    
\end{figure}

To represent a range for the types of baryonic effects on the matter power spectrum that can be expected within reasonable scenarios, we consider the 9 unique models that compose the OWLS suite of hydrodynamic simulations~\citep{2010MNRAS.402.1536S,2011MNRAS.415.3649V}. A brief summary of the physical effects that are included in each model can be found in Table~1 of~\citet{2011MNRAS.415.3649V}.
These simulations were carried out using a modified version of Gadget-3, in periodic boxes with side length $100h^{-1} {\rm Mpc}$ using $512^3$ dark matter particles and an equal number of baryonic particles. Each simulation (including one that contains only dark matter) was started with the same initial conditions, so that the ratio $P_\text{baryons+DM}/P_\text{DM-only}$ can be obtained without the sample variance inherent in the initial perturbations\footnote{Of course, these ratios will likely have some dependence on the background cosmology, but since we are only using these models as representative of the possible range of baryonic effects on the matter power spectrum, we will ignore this cosmology-dependence in this work. Furthermore, the ratio $P_\text{baryons+DM}/P_\text{DM-only}$ will itself have nonzero sample variance, by virtue of the fact that it has been measured from a finite spatial volume. We will not attempt to quantify this sample variance in this work, but merely note that future precision comparisons with hydrodynamical simulations will eventually need to take this into account.}.

\begin{figure}
\includegraphics[width=\columnwidth]{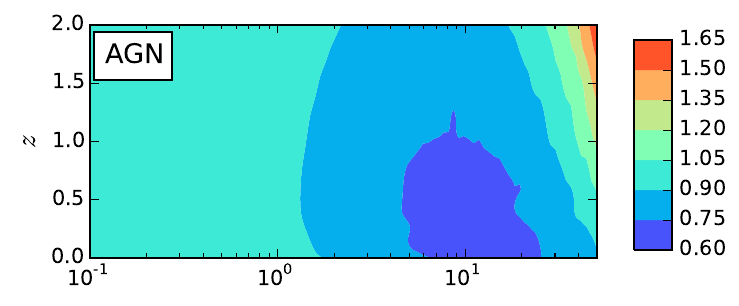}
\includegraphics[width=\columnwidth]{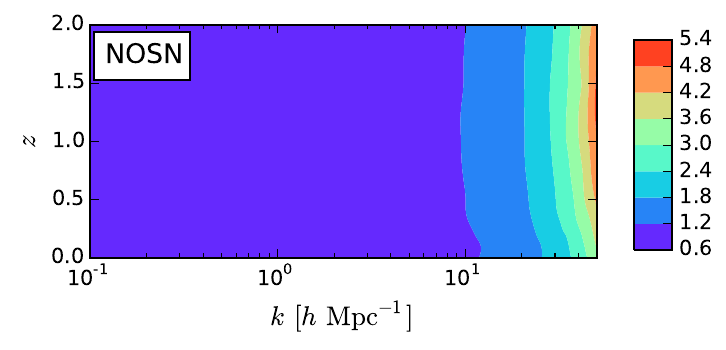}
\caption{\label{fig:baryons_contour}
As Fig.~\ref{fig:baryons_pkratio}, but plotted as a function of wavenumber and redshift, and only showing the two most extreme models: one ({\it AGN}) in which the combined effects of AGN and supernova feedback lead to a strong suppression of the power spectrum at relatively large scales, and another ({\it NOSN}) that neglects all feedback processes but contains gas cooling that enhances the power spectrum at small scales.
}    
\end{figure}

These ratios are shown in Fig.~\ref{fig:baryons_pkratio} at $z=0$ for each of the 9 models. These models can be phenomenologically classified into two groups: 
\begin{enumerate}
\item {\it AGN, DBLIMFV1618, WDENS, WML1V848}: These include strong feedback processes that drive gas out of galaxies, smoothing the total density field and thereby suppressing the power spectrum for $k\lesssim 10\invMpc$.
\item {\it REF, WML4, NOSN\_NOZCOOL, NOZCOOL, NOSN}: These contain relatively weak feedback (or none at all), with the dominant effect on the power spectrum instead being an enhancement at $k\gtrsim 5\invMpc$ arising from gas cooling that increases the central density in halos.
\end{enumerate}
In Fig.~\ref{fig:baryons_contour}, we show contour plots of the ratio $P_\text{baryons+DM}/P_\text{DM-only}$ for the two models at the extremes of this classification: {\it AGN}, which includes feedback from active galactic nuclei in addition to supernovae, and {\it NOSN}, which does not include any feedback mechanisms but includes radiative cooling.

Of the OWLS models, {\it AGN}  can be thought of as the most realistic, since the AGN feedback effects that it includes result in much better agreement with observations of low-redshift groups of galaxies than any of the other models (see~\citealt{2010MNRAS.402.1536S,2011MNRAS.415.3649V,2015MNRAS.452.3529V} for details). Some others, such as those that neglect radiative cooling processes known to be important in matching galaxy-scale observations, are decidedly unrealistic. Nevertheless we still find it useful test to test our principal-component method on the full range of scenarios.

\section{Validation of Method}
\label{sec:valopt}

The strategy of identifying the best-constrained principal components of $P(k,z)$, as described in Sec.~\ref{sec:pca}, will only be useful if these principal components can also provide a reasonable amount of information about the influence of baryonic effects on the power spectrum. In this section, we quantify the information content of the best-constrained PCs with respect to our chosen set of baryonic models, and demonstrate that a small number of these PCs provides sufficient information about the full set of models to lead to meaningful constraints.

We first define the following parametrization for each baryonic model separately, based around a free amplitude $\aowls$ that interpolates between the DM-only case ($\aowls=0$) and the exact DM+baryons power spectrum ($\aowls=1$):
\begin{align} \nn
P(k,z) &= P_\text{DM-only}(k,z)  \\
&\quad\times \lb 1 + \aowls \lp \frac{P_\text{DM+baryons}(k,z)}{P_\text{DM-only}(k,z)} - 1 \rp \rb\ .
\label{eq:aowls_def}
\end{align}
If we were only interested in a single baryonic model, we could imagine using Eq.~\eqref{eq:aowls_def} as our model for baryonic effects, and attempting to constrain $\aowls$ along with the other cosmological and systematics parameters. Alternatively, we can parametrize baryonic effects using the amplitudes~$\{\alpha_a\}$ of the $N$ most important PCs. To do so, we can solve Eq.~\eqref{eq:alphaa_def} for $\Delta_\mu$ at each $\mu\equiv(k_i,z_j)$, using only up to mode $N$ on the right hand side, and then smoothly interpolate between these discrete values of $\Delta_\mu$ to construct a function $\Delta_N(k,z)$, which will depend on the values $\{\alpha_1,\cdots,\alpha_N\}$. The power spectrum is then given by
\beq
P(k,z) = P_\text{DM-only}(k,z) \lb 1 + \Delta_N(k,z)  \rb\ .
\eeq

An intuitive way to determine how much information the PCs provide about different baryonic models is to relate constraints on the amplitudes $\{\alpha_a\}$ to a constraint on~$\aowls$. If we treat each amplitude as providing an independent ``measurement" of~$\aowls$ and apply inverse-variance weighting to these ``measurements," we obtain\footnote{The right-hand side of Eq.~\eqref{eq:sig2aowls} is also the square of the cumulative signal-to-noise of the first $N$ PCs, and therefore can alternatively be interpreted without the need of the $\aowls$ parametrization from Eq.~\eqref{eq:aowls_def}.}
\beq
\frac{1}{\sigma^2(\aowls)} = \sum_{a=1}^N \frac{1}{\sigma_a^2} \lp \frac{\d\alpha_a}{\d\aowls} \rp^2
	= \sum_{a=1}^N \frac{ \lp \alphaowls \rp^2 }{\sigma_a^2} \ ,
\label{eq:sig2aowls}
\eeq
where $\alphaowls$ is the projection of the specific baryonic model onto the chosen PCs:
\beq
\alphaowls = \sum_\mu \beta_{a\mu} \Deltaowls\ .
\eeq
(The second equality in Eq.~\eqref{eq:sig2aowls} follows from the fact that $\Deltaowls$, and therefore $\alphaowls$, is linear in $\aowls$.) 

\begin{figure}
\includegraphics[width=\columnwidth]{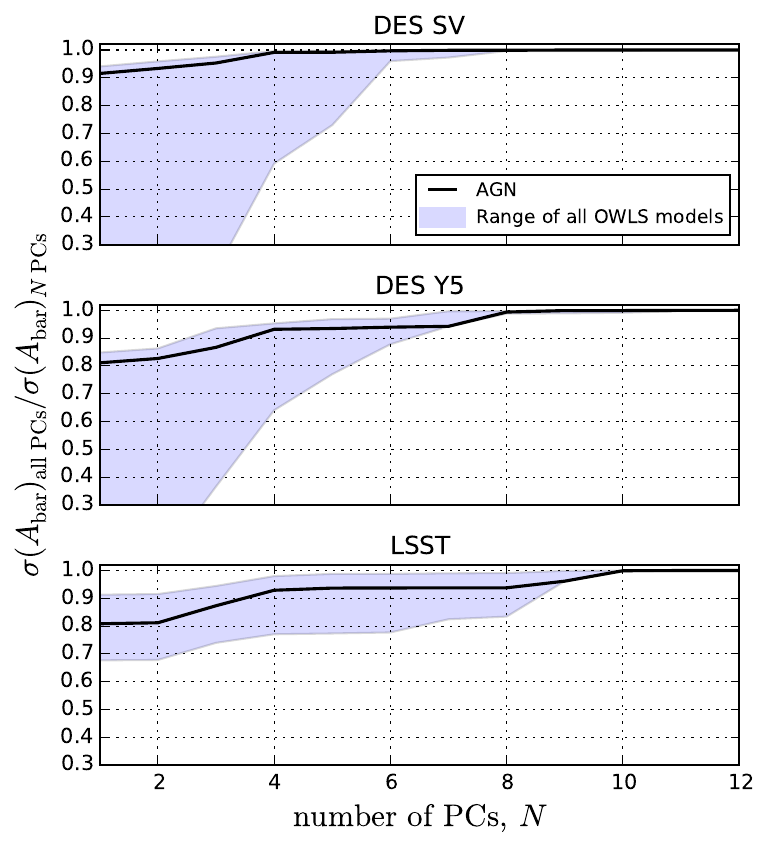}
\caption{\label{fig:sigma_aowls_nmodes_surveys}
Ratio of the effective one-sigma constraint on $\aowls$ (defined in Eq.~\eqref{eq:aowls_def}), $\sigma(\aowls)$ from using all available PCs with that from using only the $N$ best-constrained PCs for the three surveys specified above each panel. For each set of PCs, the effective constraint on $\aowls$ is calculated using Eq.~\eqref{eq:sig2aowls}. The black curve corresponds to the OWLS {\it AGN} model, while the blue band represents the range spanned by all OWLS models. This represents the fraction of the full survey's information that is retained by the first $N$ PCs of $P(k,z)$. Other Stage IV surveys give results comparable to LSST, and therefore we conclude that for all surveys we consider, the first $\sim$9 PCs contain 90\% of the total constraining power on our test set of baryonic models.
}    
\end{figure}

When {\it all} available PCs are used in Eq.~\eqref{eq:sig2aowls} to determine $\sigma(\aowls)$, the result is very close to the output of forecasts in which we constrain $\aowls$ directly for each baryonic model in our testing set. However, it is sufficient to use a much smaller number of PCs to retain a comparable amount of constraining power. In Fig.~\ref{fig:sigma_aowls_nmodes_surveys}, we check how this constraining power scales with the number $N$ of the most important PCs we retain, by plotting the ratio of $\sigma(\aowls)$ obtained from all PCs and that from the first $N$ PCs. If we wish to retain 90\% of the constraining power of the full set of PCs, we may only keep the first 9 PCs in our analysis. This number is an upper bound across all surveys we have investigated, including those not shown in Fig.~\ref{fig:sigma_aowls_nmodes_surveys}, and can be considered as a validation of the usefulness of the PCA approach.

Before moving on, we remind the reader that the intended application of the PCA method to data would make use of the joint posteriors on the amplitudes of all relevant PCs, rather than compressing this information into a single $\aowls$ parameter for each specific model. We have used the $\aowls$ parametrization merely as a convenient way to assess and quantify the performance of the PCA method in various cases.

In the next section, we will perform forecasts for the constraining power of the first 9 PCs of various surveys. We use the same number of PCs for each survey in order to make meaningful comparisons, but in principle, a given survey might require even fewer PCs to achieve meaningful constraints. For example, the top panel of Fig.~\ref{fig:sigma_aowls_nmodes_surveys} shows that for cosmic shear measurements comparable to those of DES SV, only the first 6 modes are required to capture 90\% of the full amount of information about the impact of baryons of the matter power spectrum. Optimization of the number of PCs in other cases is best left to future, survey-specific analyses.

\section{Forecasts}
\label{sec:forecasts}

\subsection{Performance of various surveys}
\label{sec:forecasts-fom}

In order to quantify the constraining power of different surveys with respect to generic baryonic effects on the matter power spectrum, we define a figure of merit $\fom$ as the reciprocal of the geometric mean of the 68\% confidence intervals for the most important 9 PCs for each survey:
\beq
\label{eq:fom}
\fom \equiv \lb \prod_{a=1}^{9}  \sigma_a \rb^{-1/9}\ .
\eeq
Since the PCs are uncorrelated by construction, the product $\prod_{a=1}^{9} \sigma_a$ is proportional to the  volume of the hyperellipsoid corresponding to the joint 68\% confidence region of all PCs, but this product differs by many orders of magnitude for different surveys. On the other hand, the quantity in Eq.~\eqref{eq:fom} efficiently captures a kind of average of the constraints on the first 9 modes; from Eq.~\eqref{eq:sig2aowls}, we then find that the effective constraints on a generic model for baryonic effects (as parametrized by $\sigma(\aowls)$) will scale proportionally to~$\fom$.

\begin{figure}
\includegraphics[width=\columnwidth]{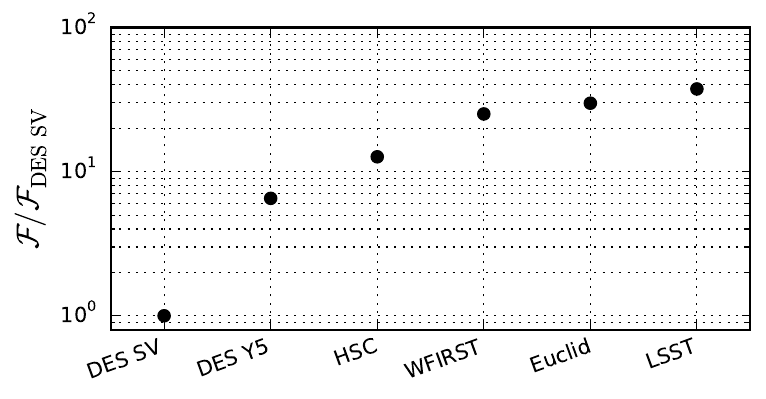}
\caption{\label{fig:fom}
Figure of merit $\fom$ (defined in Eq.~\eqref{eq:fom}) for constraints on baryonic models, normalized to the value for DES SV, for forecasts performed with mock $\xi_\pm$ measurements down to $\theta_{\rm min}=2'$. We find that both HSC and the Y5 shear measurements of cosmic shear from DES will improve $\fom$ by roughly an order of magnitude, implying a similar potential improvement in our information about baryonic effects on the matter power spectrum, while Stage IV surveys like LSST will improve on that by a further factor of a few.
}    
\end{figure}

Fig.~\ref{fig:fom} presents the figure of merit for each survey we consider in this work, for forecasts using $\xi_\pm$ measurements down to $\theta_{\rm min}=2'$, normalized to the figure of merit for DES SV to emphasize the improvement of future shear measurements over currently available data. Fig.~\ref{fig:fom} indicates that HSC and the Y5 shear measurements from DES will improve on current data by roughly an order of magnitude, while upcoming Stage IV surveys will only improve on that by a factor of a few. This can be understood intuitively by making use of the following rough scalings:
\begin{itemize}
\item $\mathcal{F} \propto f_{\rm sky}^{0.5}$, since the leading terms in the data covariance matrix scale like $f_{\rm sky}^{-1}$.
\item Because of shot noise in the shear correlation function, $\mathcal{F} \propto (n\sigma_\epsilon^{-2})^{0.6}$ (this power has been empirically determined by varying $n$ in forecasts for various surveys). If shot noise completely dominated the covariance matrix at all scales, we would expect $\mathcal{F} \propto (n\sigma_\epsilon^{-2})^1$, but the balance between the cosmic variance and shot noise terms in Eq.~\eqref{eq:clplussn} softens the dependence to what we quote here.
\item On average, the figure of merit scales with the number of redshift bins like $\mathcal{F} \propto n_{\rm bins}^{0.4}$. This power is steeper for surveys with broader redshift distributions, since in those cases, narrower redshift bins contain more information than for surveys that primarily probe lower redshifts.
\end{itemize}
Therefore, it is essentially the improvement in the expected number of galaxies with measured shapes (i.e.\ $f_{\rm sky} \times n$) that drives the large improvement of Stage III surveys over current data, as compared to the somewhat more modest improvement afforded by Stage IV surveys.

\begin{figure*}
\includegraphics[width=\textwidth]{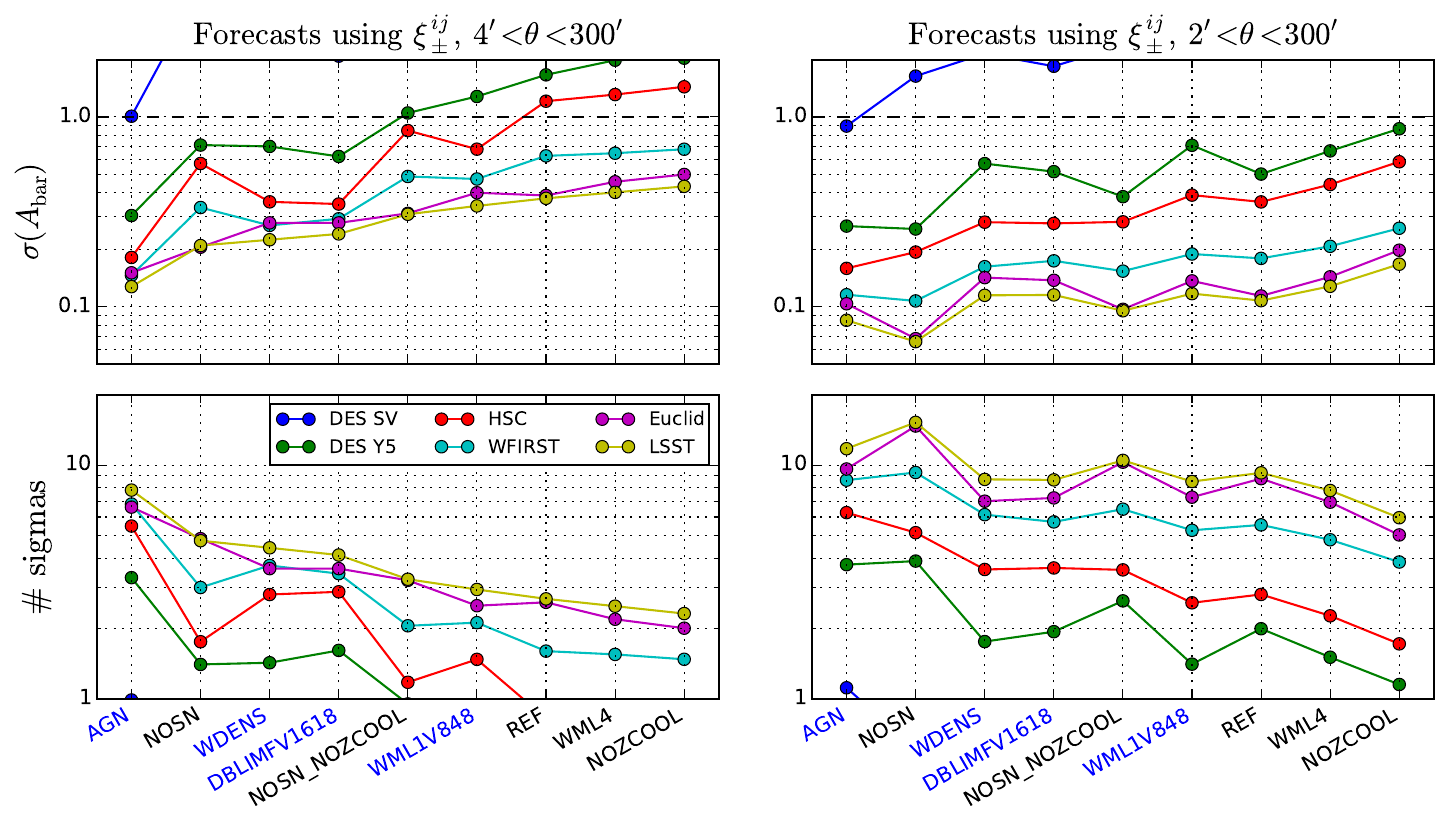}
\caption{\label{fig:sigma_aowls_absolute}
{\it Top panels:} Expected one-sigma constraints on $\aowls$ (see Eq.~\eqref{eq:aowls_def}) for each survey and baryonic model we consider, when the first 9 PCs for each survey are used in the analysis. {\it Bottom panels:} Statistical significance at which each survey could rule out each model, again using the PCA approach with the amplitudes of the first 9 PCs included as free parameters in the analysis. Using measurements of $\xi_\pm^{ij}$ over the range $4'<\theta<300'$ ({\it left panels}), the constraints are strongest on models including prescriptions for feedback or other effects that affect the power spectrum at relatively large scales, e.g.~$k\lesssim1\invMpc$ at $z=0$ ({\it blue labels}). When $\theta_{\rm min}=2'$ is used instead ({\it right panels}), models with large deviations from the DM-only power spectrum on small scales (typically due to cooling effects on the centers of halos; {\it black labels}) are much better constrained due to the inclusion of the additional small-scale information.
}
\end{figure*}

\begin{figure}
\includegraphics[width=\columnwidth]{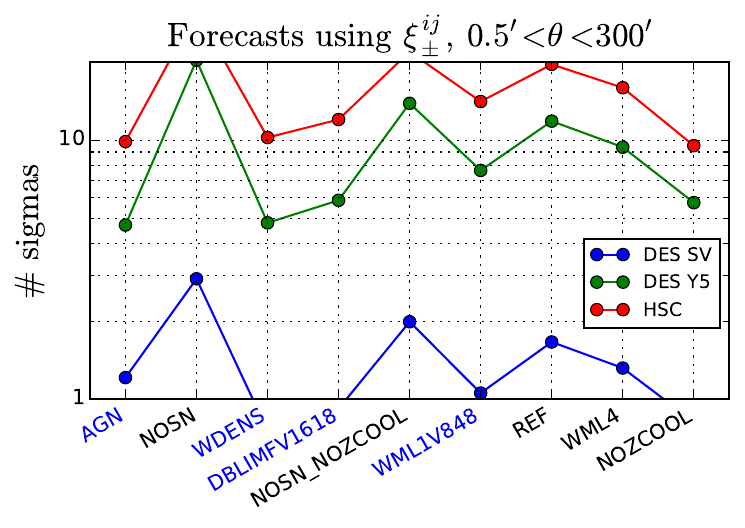}
\caption{\label{fig:sigma_aowls_absolute_xi_thmin0p5}
Same as the bottom panels of Fig.~\ref{fig:sigma_aowls_absolute}, but assuming that measurements of $\xi_\pm$ can be used down to $\theta_{\rm min}=0.5'$. The gain in constraining power is significant (and also robust to the variations in systematics discussed in Sec.~\ref{sec:sys-priors}), with all other surveys we have considered (not shown) providing better than 20$\sigma$ constraints across the entire test set of baryonic models. Therefore, the coming generation of cosmic shear measurements could potentially be very informative with regards to the implementation of certain baryonic phenomena in simulations.
}    
\end{figure}

\begin{figure}
\includegraphics[width=\columnwidth]{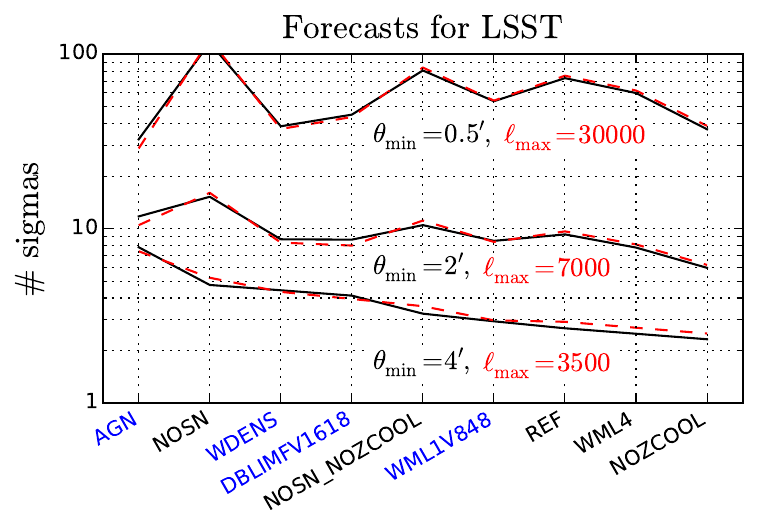}
\caption{\label{fig:xi_vs_cl}
Same as the bottom panels of Fig.~\ref{fig:sigma_aowls_absolute}, but comparing LSST forecasts using the real-space correlation function $\xi_\pm$ ({\it black solid lines}) or the angular power spectrum $C_\ell$ ({\it red dashed lines}) with different values of $\theta_{\rm min}$ and $\ell_{\rm max}$ as indicated in the figure. For each value of $\theta_{\rm min}$, we find that $\ell_{\rm max}$ can be chosen to provide very similar constraints on our test set of baryonic models, following the relationship $\theta_{\rm min} \simeq 1.3\pi/\ell_{\rm max}$.
}    
\end{figure}

\subsection{Expected constraints on specific baryonic models}

Beyond the model-independent approach of Sec.~\ref{sec:forecasts-fom}, we can also ask how our method can provide information about specific baryonic models when applied to different surveys. Specifically, we can use Eq.~\eqref{eq:sig2aowls} to translate the expected constraints on the 9 most important PCs into constraints on $\aowls$ for each of the models in Sec.~\ref{sec:baryon_models}. We can further translate these constraints into the statistical significance at which each model could be ruled out by a given survey (or, alternatively, the significance at which the DM-only power spectrum can be ruled out).

The top left panel of Fig.~\ref{fig:sigma_aowls_absolute} shows the resulting values for $\sigma(\aowls)$ for each survey and baryonic model, in the case where the shear correlation functions $\xi_\pm^{ij}$ have been measured down to $\theta_{\rm min}=4'$, while the bottom left panel shows the corresponding statistical significance of these constraints, given in terms of the ``number of sigmas" at which the model could be distinguished from the DM-only case. Unsurprisingly, the models that can be best constrained by the most powerful surveys are {\it AGN}, whose implementation of feedback has dramatic effects at large scales\footnote{Other more recent hydrodynamical simulations, such as Illustris~\citep{2014Natur.509..177V}, have even stronger effects on the power spectrum at large scales, and therefore can likely be constrained at least at the level of the {\it AGN} model we examine here.}, and {\it NOSN}, which lacks any feedback processes but implies a large enhancement at small scales due to cool gas altering the cores of halos. Other models with significant feedback or strong winds, such as {\it WDENS}, are the next-best constrained, in some cases even better than {\it NOSN} due to shape noise and other factors that reduce the capabilities of some surveys to probe the smallest scales. 

When we consider measurements of $\xi_\pm$ down to $\theta_{\rm min}=2'$, shown in the right panels of Fig.~\ref{fig:sigma_aowls_absolute},  the corresponding increase in small-scale information serves to improve the constraining power by roughly a factor of two in many cases\footnote{We have also checked the case where $\theta_{\rm min}$ is set to $2'$ for $\xi_+$ and $\theta_{\rm min}=10'$ for $\xi_-$, since different scale cuts are sometimes used for $\xi_+$ and $\xi_-$ due to the increased vulnerability of $\xi_-$ to small-systematics as compared to the $\xi_+$. The results in this case are very similar to what we obtain using $\theta_{\rm min}=4'$ for both $\xi_+$ and $\xi_-$.}. The improvement is more modest for models where feedback reduces the impact of cooling on the clustering at the smallest scales, and in some cases, the slight shift in the angular coordinates of the measured $\xi_\pm$ data points causes a very slight degradation in the constraints.

Given the importance of small scales for this analysis, we have also investigated the case where measurements of $\xi_\pm$ can be used down to $\theta_{\rm min}=0.5'$. The results for DES SV-like or Stage III datasets are shown in Fig.~\ref{fig:sigma_aowls_absolute_xi_thmin0p5}; for all other surveys we consider, the constraints on all baryonic models are better than 20$\sigma$. This conclusion is insensitive to the priors on systematic effects that we will return to in Sec.~\ref{sec:sys-priors}, but of course relies on sufficient control of all observational issues that could affect the usability of measurements at these scales. Nevertheless, it seems likely that data from Stage III surveys, or even partial data releases from these surveys,  will be informative with respect to certain behaviors of baryons on the scales probed by cosmic shear measurements.

We have presented our results in the framework of distinguishing each model from the DM-only case, but with the high level of statistical significance that will be afforded by future surveys, it is likely that these measurements will be able to distinguish between different scenarios themselves. This can be accomplished by examining the joint posteriors on the amplitudes of each PC, and then comparing the projections of individual models onto these PCs. This procedure can in principle be applied to any model or simulation for which the total matter power spectrum is known, and may also be useful in constraining models containing their own continuous parameters.

Similar conclusions apply to cases where two-point statistics other than the correlation functions $\xi_\pm$ are used in the analysis. In Fig.~\ref{fig:xi_vs_cl}, we compare forecasts for LSST performed with either $\xi_\pm$ or the angular power spectrum $C_\ell$, with $\ell_{\rm max}$ chosen to reproduce as closely as possible the results from $\xi_\pm$ for each value of $\theta_{\rm min}$ we have previously considered. It is to be expected that $\theta_{\rm min}$ scales roughly as $\pi/\ell_{\rm max}$, and our explicit comparisons bear out this expectation, with $\theta_{\rm min} \simeq 1.3\pi/\ell_{\rm max}$. Specifically, we find that the $\xi_\pm$ forecasts with $\theta_{\rm min} = \{4',2',0.5'\}$ match quite closely with $C_\ell$ forecasts with $\ell_{\rm max}=\{3500,7000,30000\}$, respectively.

\subsection{Impact of systematics at small scales}
\label{sec:sys-priors}

\begin{figure*}
\includegraphics[width=1.0\textwidth]{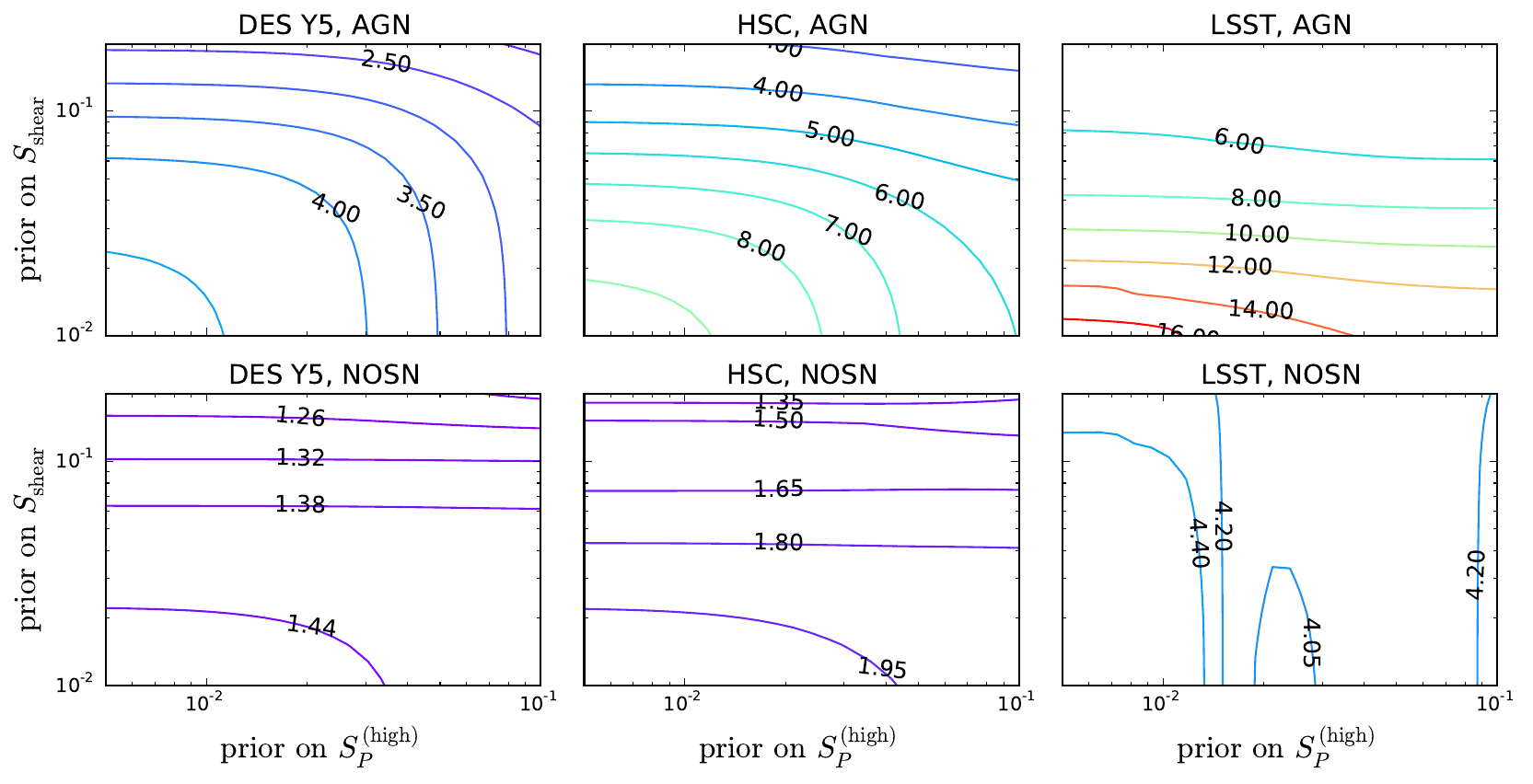}
\caption{\label{fig:sys-xi-tm4}
For $\xi_\pm$ measurements with $\theta_{\rm min}=4'$, each panel shows the statistical significance (in ``number of sigmas", as in Figs.~\ref{fig:sigma_aowls_absolute}-\ref{fig:xi_vs_cl}) at which the specified baryonic model can be ruled out by the specified survey, as a function of the assumed level of two types of systematic uncertainty in the theoretical modeling. The parameter $S_\text{shear}$ denotes the uncertainty in the handling of shear-specific systematics (such as source-lens clustering) in the modeling of the shear correlation function, assumed to scale like $\Delta C_\ell \propto \ell^{0.5}$, while $S_P^\text{(high)}$ denotes the uncertainty in the DM-only power spectrum for $k<0.5\invMpc$ (see Sec.~\ref{sec:systematics} for details). Accuracy in both aspects of the modeling is more important for {\it AGN}-like models in which feedback affects the power spectrum on relatively large scales, while for cooling-dominated models, $S_P^\text{(high)}$ is less relevant. Meanwhile, Stage IV surveys will be sufficiently powerful that relatively large ($\sim$10\%) uncertainties in the modeling will not appreciably degrade the constraints.
}    
\end{figure*}

Of the possible systematics we describe in Sec.~\ref{sec:systematics}, two of the more uncertain are the accuracy of the modeling of the DM-only power spectrum (including the effects of massive neutrinos) and the level of shear-specific systematics (e.g.\ lensing bias) that could be important when interpreting measurements at small scales. Our fiducial choice for the main results of the paper has been a 5\% uncertainty in the DM-only power spectrum for $k>0.5\invMpc$ (our $S_P^{\rm (high)}$ parameter) and a 5\% error budget for shear-specific systematics at $\ell=10^4$ (our $S_{\rm shear}$ parameter), with a profile scaling like $\Delta C_\ell \propto \ell^{0.5}$. 

\begin{figure*}
\includegraphics[width=1.0\textwidth]{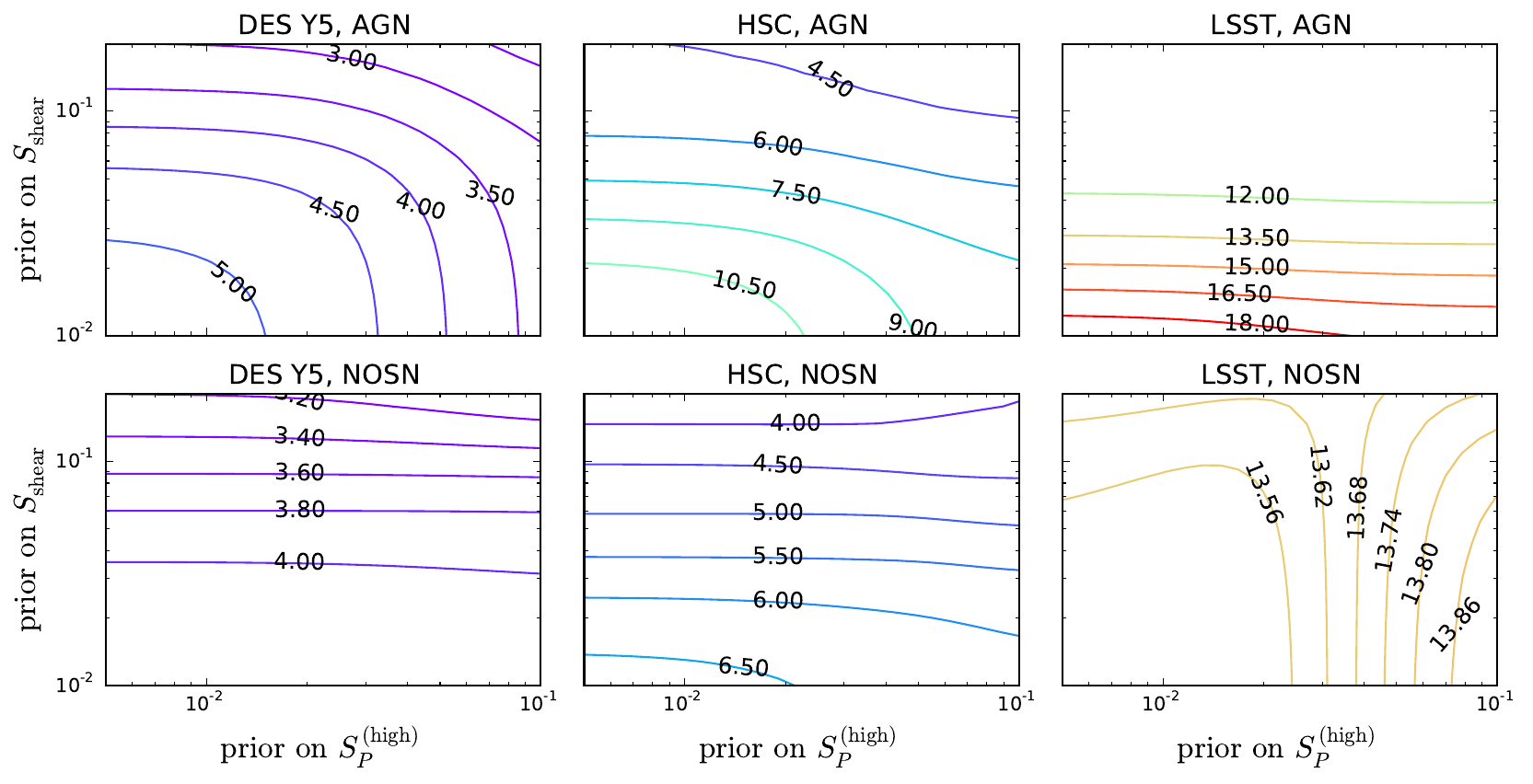}
\caption{\label{fig:sys-xi-tm2}
As Fig.~\ref{fig:sys-xi-tm4}, but corresponding to $\xi_\pm$ measurements with $\theta_{\rm min}=2'$ instead of $4'$.
}    
\end{figure*}

In Figs.~\ref{fig:sys-xi-tm4} and~\ref{fig:sys-xi-tm2}, we display representative variations in our forecasts when the priors on $S_P^{\rm (high)}$ and $S_{\rm shear}$ are varied away from their fiducial values, for the forecasts with $\theta_{\rm min}=4'$ and $2'$ respectively. For the {\it AGN} model and others where feedback affects the power spectrum on large scales, the accuracy in the modeling of the DM-only power spectrum is an important determinant of the overall constraining power, provided that the other shear systematics are controlled at a reasonable level. On the other hand, for models for which the constraining power is concentrated at smaller scales, the effects of baryons on the power spectrum at these scales are generally so strong that  knowledge of the DM-only power spectrum to within 10\% will be sufficient to detect their presence from shear measurements alone.

Essentially the same conclusion holds for the prior on $S_{\rm shear}$: it has a relatively smaller impact on cooling-dominated models than on feedback-dominated scenarios. The increased precision of Stage IV surveys like LSST actually translates into {\it less} stringent requirements on the modeling of the DM-only power spectrum for {\it AGN}-like models, although the main science goals of these surveys (e.g.\ searches for new cosmological physics) will of course impose their own stricter requirements. In fact, for cooling-dominated models, we find that the constraints from Stage IV surveys are fairly insensitive to the systematics we vary in this section; measurements with $\theta_{\rm min} \lesssim 3'$ will be capable of distinguishing our entire test set of baryonic models from the DM-only case, even with $\sim$10\% uncertainties in the theoretical modeling on the relevant scales.

\section{Conclusions}
\label{sec:conclusions}
In this paper, we have assessed the ability of the two-point statistics of cosmic shear to act as a probe of the physics of galaxy formation, which is expected to impact the matter power spectrum. We have done so using a method that parameterizes the effects of baryons on the matter power spectrum in terms of a set of principal components (PCs). For a given survey, the method automatically classifies these PCs in terms of the expected constraints on each one, allowing for poorly-constrained PCs to be discarded from an analysis without a significant loss of information. These PCs are determined by the properties of a given survey, and are independent of the implementations of baryonic physics one might be interested in testing.

We have performed forecasts for the application of this method to a variety of surveys, ranging from those that have already been completed (the Dark Energy Survey Science Verification run) to Stage III (DES Y5, HSC) and Stage IV (WFIRST, LSST, Euclid) surveys. We have used the OWLS simulation suite, composed of nine different implementations of baryonic effects plus a reference dark matter-only run, to test how the (model-independent) output of the method translates into constraints on a range of specific models. 

We find that surveys operating now and planned in the near future have significant power to constrain the impact of galaxy formation on the matter power spectrum, with the potential to strongly constrain or rule out a majority of the models considered here. 
(Indeed, in~\citealt{2015MNRAS.450.1212H}, constraints were already obtained from CFHTLenS data, with similar signal-to-noise to the DES SV curve in Fig.~\ref{fig:sigma_aowls_absolute_xi_thmin0p5}, although using a different method than what we present here.)
We emphasize that cosmic shear is a very clean signal to model: the connection between the observed correlations of galaxy shapes and the underlying matter clustering is completely determined by a weak-field calculation in general relativity, and does not entail any of the astrophysical calibration or selection issues involved in many other probes of galaxy formation physics. 

Our main results may be summarized as follows:
\begin{itemize}
\item For all surveys we consider, we find that 90\% of the constraining power with respect to baryonic effects on the matter power spectrum is encapsulated in no more than nine PCs, implying a likelihood analysis that makes use of this PCA method would need to add no more than nine additional parameters.
\item From these nine PCs, one can define a figure of merit as the reciprocal of the geometric mean of the expected one-sigma constraints on the corresponding amplitudes. By comparing this figure of merit for different surveys, we find that the constraints from Stage III surveys will improve on those from currently available shear measurements by roughly an order of magnitude, while Stage IV surveys will provide further improvements of a factor of a few. We argue that it is chiefly the number of galaxies with measured shapes that drives this improvement.
\item We find that the ultimate power of our method to constrain baryonic effects depends strongly on the minimum scale $\theta_{\rm min}$ at which measurements of the shear correlation functions  $\xi_\pm$ can be used. In particular, if $\theta_{\rm min}$ can be pushed down to less than one arcminute, Stage III surveys should be able to rule out (at more than five-sigma confidence) the majority of the baryonic models we consider. We reach this conclusion after marginalizing over uncertainties in neutrino mass, photometric redshift distributions, shear calibration, and theoretical modeling of the power spectrum.
\end{itemize}

In our analysis, we have assumed that the background cosmology has already been mostly fixed by other measurements, such as temperature fluctuations in the cosmic microwave background. For our fiducial cosmology, we have chosen the latest best-fit model from Planck~\citep{2015arXiv150201589P}, marginalizing over their one-sigma uncertainties. In applying our method to data, care must be taken in choosing a fiducial cosmology that is consistent with the dataset being employed (accounting, for example, for the current tensions between cosmological constraints from weak lensing and the CMB), to avoid a wrong choice of cosmology biasing the constraints on baryonic effects. While we have not investigated the size of this possible bias in detail, it could easily be investigated in a future analysis simply by varying the choice of cosmology and examining the resulting change in the constraints on the PCs.

The primary product of a likelihood analysis making use of this method would be posterior constraints on the amplitudes of the most important PCs for that survey, with the PCs determined by a similar Fisher matrix procedure to what we have used here. The power spectrum from a given baryonic model can then be projected onto these PCs, and the resulting amplitudes can be compared with the constraints, without the need for an separate likelihood analysis corresponding to every model of interest. In this way, the results of a likelihood analysis including the PCs will act as a resource that can be used to assess the realism of currently-existing models or hydrodynamical simulations, and could also be used to guide the development of new models or simulations.

Furthermore, there currently exist models for baryonic effects on clustering that contain parameters connected to specific physical processes, such as halo expansion or gas ejection (e.g.~\citealt{2015MNRAS.454.1958M,2015JCAP...12..049S}). Constraints on the PCs from our method could also be mapped onto constraints on such parameters, and could therefore help to inform our knowledge about the related processes, in a more detailed manner than a wholesale acceptance or rejection of an individual model.

Finally, we note that even stronger constraints can likely be obtained by performing this kind of analysis on the combination of cosmic shear and other cosmological statistics. These include, for example, the cluster--mass correlation function, cluster Sunyaev-Zeldovich profiles, correlations between weak lensing convergence and thermal Sunyaev-Zel'dovich maps~\citep{2014PhRvD..89b3508V,2015JCAP...09..046M,2015JCAP...10..047H}, and galaxy--galaxy lensing.  These may have significant additional power but introduce extra complexity in the modeling due for example to the need for a model for the cluster--halo connection in the first two cases and a bias model in the fourth case. Regardless, we have shown that, if sub-arcminute measurements can be robustly made, the connection between cosmic shear and the overall clustering of matter can be exploited to turn cosmic shear into an important probe of the physics of galaxy formation.

\section*{Acknowledgements}

This work received partial support from the U.S.\ Department of Energy under contract number DE-AC02-76SF00515. 
We thank Tim Eifler, Andrew Hearin, Dragan Huterer, Elisabeth Krause, Niall MacCrann, Patrick Simon, and Michael Troxel for useful discussions and comments. We also thank Elisabeth Krause for providing code for computing the covariance matrices used in this work, and Stuart Marshall and the SLAC computing staff for essential computing support. This work has made extensive use of results from the OWLS simulation suite, and we thank the authors for making these data available from~\href{http://vd11.strw.leidenuniv.nl/}{http://vd11.strw.leidenuniv.nl/}.
This work also made use of the CosmoSIS software package, available from~\href{http://bitbucket.org/joezuntz/cosmosis}{http://bitbucket.org/joezuntz/cosmosis}.  We thank Joe Zuntz and the CosmoSIS team for making this code available.

\bibliographystyle{mnras_sjf}
\bibliography{references} 

\begin{thebibliography}{}
\makeatletter
\relax
\def\mn@urlcharsother{\let\do\@makeother \do\$\do\&\do\#\do\^\do\_\do\%\do\~}
\def\mn@doi{\begingroup\mn@urlcharsother \@ifnextchar [ {\mn@doi@}
  {\mn@doi@[]}}
\def\mn@doi@[#1]#2{\def\@tempa{#1}\ifx\@tempa\@empty \href
  {http://dx.doi.org/#2} {doi:#2}\else \href {http://dx.doi.org/#2} {#1}\fi
  \endgroup}
\def\mn@eprint#1#2{\mn@eprint@#1:#2::\@nil}
\def\mn@eprint@arXiv#1{\href {http://arxiv.org/abs/#1} {{\tt arXiv:#1}}}
\def\mn@eprint@dblp#1{\href {http://dblp.uni-trier.de/rec/bibtex/#1.xml}
  {dblp:#1}}
\def\mn@eprint@#1:#2:#3:#4\@nil{\def\@tempa {#1}\def\@tempb {#2}\def\@tempc
  {#3}\ifx \@tempc \@empty \let \@tempc \@tempb \let \@tempb \@tempa \fi \ifx
  \@tempb \@empty \def\@tempb {arXiv}\fi \@ifundefined
  {mn@eprint@\@tempb}{\@tempb:\@tempc}{\expandafter \expandafter \csname
  mn@eprint@\@tempb\endcsname \expandafter{\@tempc}}}

\bibitem[\protect\citeauthoryear{{Albrecht} et~al.,}{{Albrecht}
  et~al.}{2009}]{2009arXiv0901.0721A}
{Albrecht} A.,  et~al., 2009, \href
  {http://adsabs.harvard.edu/abs/2009arXiv0901.0721A} {} \mn@eprint {arXiv}
  {0901.0721}

\bibitem[\protect\citeauthoryear{{Bacon}, {Refregier}  \& {Ellis}}{{Bacon}
  et~al.}{2000}]{2000MNRAS.318..625B}
{Bacon} D.~J.,  {Refregier} A.~R.,   {Ellis} R.~S.,  2000, \mn@doi [\mnras]
  {10.1046/j.1365-8711.2000.03851.x}, \href
  {http://adsabs.harvard.edu/abs/2000MNRAS.318..625B} {318, 625},  \mn@eprint
  {} {astro-ph/0003008}

\bibitem[\protect\citeauthoryear{{Becker} et~al.,}{{Becker}
  et~al.}{2015}]{2015arXiv150705598B}
{Becker} M.~R.,  et~al., 2015, \href
  {http://adsabs.harvard.edu/abs/2015arXiv150705598B} {} \mn@eprint {arXiv}
  {1507.05598}

\bibitem[\protect\citeauthoryear{{Bernstein}}{{Bernstein}}{2009}]{2009ApJ...695..652B}
{Bernstein} G.~M.,  2009, \mn@doi [\apj] {10.1088/0004-637X/695/1/652}, \href
  {http://adsabs.harvard.edu/abs/2009ApJ...695..652B} {695, 652},  \mn@eprint
  {arXiv} {0808.3400}

\bibitem[\protect\citeauthoryear{{Bielefeld}, {Huterer}  \&
  {Linder}}{{Bielefeld} et~al.}{2015}]{2015JCAP...05..023B}
{Bielefeld} J.,  {Huterer} D.,   {Linder} E.~V.,  2015, \mn@doi [\jcap]
  {10.1088/1475-7516/2015/05/023}, \href
  {http://adsabs.harvard.edu/abs/2015JCAP...05..023B} {5, 023},  \mn@eprint
  {arXiv} {1411.3725}

\bibitem[\protect\citeauthoryear{{Bird}, {Viel}  \& {Haehnelt}}{{Bird}
  et~al.}{2012}]{2012MNRAS.420.2551B}
{Bird} S.,  {Viel} M.,   {Haehnelt} M.~G.,  2012, \mn@doi [\mnras]
  {10.1111/j.1365-2966.2011.20222.x}, \href
  {http://adsabs.harvard.edu/abs/2012MNRAS.420.2551B} {420, 2551},  \mn@eprint
  {arXiv} {1109.4416}

\bibitem[\protect\citeauthoryear{{Bonnett} et~al.,}{{Bonnett}
  et~al.}{2015}]{2015arXiv150705909B}
{Bonnett} C.,  et~al., 2015, \href
  {http://adsabs.harvard.edu/abs/2015arXiv150705909B} {} \mn@eprint {arXiv}
  {1507.05909}

\bibitem[\protect\citeauthoryear{{Bridle} \& {King}}{{Bridle} \&
  {King}}{2007}]{2007NJPh....9..444B}
{Bridle} S.,  {King} L.,  2007, \mn@doi [New Journal of Physics]
  {10.1088/1367-2630/9/12/444}, \href
  {http://adsabs.harvard.edu/abs/2007NJPh....9..444B} {9, 444},  \mn@eprint
  {arXiv} {0705.0166}

\bibitem[\protect\citeauthoryear{{Dvorkin} \& {Hu}}{{Dvorkin} \&
  {Hu}}{2010}]{2010PhRvD..82d3513D}
{Dvorkin} C.,  {Hu} W.,  2010, \mn@doi [\prd] {10.1103/PhysRevD.82.043513},
  \href {http://adsabs.harvard.edu/abs/2010PhRvD..82d3513D} {82, 043513},
  \mn@eprint {arXiv} {1007.0215}

\bibitem[\protect\citeauthoryear{{Eifler}, {Schneider}  \& {Hartlap}}{{Eifler}
  et~al.}{2009}]{2009A&A...502..721E}
{Eifler} T.,  {Schneider} P.,   {Hartlap} J.,  2009, \mn@doi [\aap]
  {10.1051/0004-6361/200811276}, \href
  {http://adsabs.harvard.edu/abs/2009A%26A...502..721E} {502, 721},  \mn@eprint
  {arXiv} {0810.4254}

\bibitem[\protect\citeauthoryear{{Eifler}, {Krause}, {Dodelson}, {Zentner},
  {Hearin}  \& {Gnedin}}{{Eifler} et~al.}{2015}]{2015MNRAS.454.2451E}
{Eifler} T.,  {Krause} E.,  {Dodelson} S.,  {Zentner} A.~R.,  {Hearin} A.~P.,
  {Gnedin} N.~Y.,  2015, \mn@doi [\mnras] {10.1093/mnras/stv2000}, \href
  {http://adsabs.harvard.edu/abs/2015MNRAS.454.2451E} {454, 2451},  \mn@eprint
  {arXiv} {1405.7423}

\bibitem[\protect\citeauthoryear{{Foreman}, {Perrier}  \& {Senatore}}{{Foreman}
  et~al.}{2016}]{2015arXiv150705326F}
{Foreman} S.,  {Perrier} H.,   {Senatore} L.,  2016, \mn@doi [\jcap]
  {10.1088/1475-7516/2016/05/027}, \href
  {http://adsabs.harvard.edu/abs/2015arXiv150705326F} {5, 027},  \mn@eprint
  {arXiv} {1507.05326}

\bibitem[\protect\citeauthoryear{{Harnois-D{\'e}raps}, {van Waerbeke}, {Viola}
  \& {Heymans}}{{Harnois-D{\'e}raps} et~al.}{2015}]{2015MNRAS.450.1212H}
{Harnois-D{\'e}raps} J.,  {van Waerbeke} L.,  {Viola} M.,   {Heymans} C.,
  2015, \mn@doi [\mnras] {10.1093/mnras/stv646}, \href
  {http://ads.nao.ac.jp/abs/2015MNRAS.450.1212H} {450, 1212},  \mn@eprint
  {arXiv} {1407.4301}

\bibitem[\protect\citeauthoryear{{Hartlap}, {Hilbert}, {Schneider}  \&
  {Hildebrandt}}{{Hartlap} et~al.}{2011}]{2011A&A...528A..51H}
{Hartlap} J.,  {Hilbert} S.,  {Schneider} P.,   {Hildebrandt} H.,  2011,
  \mn@doi [\aap] {10.1051/0004-6361/201015850}, \href
  {http://adsabs.harvard.edu/abs/2011A%26A...528A..51H} {528, A51},  \mn@eprint
  {arXiv} {1010.0010}

\bibitem[\protect\citeauthoryear{{Hearin}, {Zentner}  \& {Ma}}{{Hearin}
  et~al.}{2012}]{2012JCAP...04..034H}
{Hearin} A.~P.,  {Zentner} A.~R.,   {Ma} Z.,  2012, \mn@doi [\jcap]
  {10.1088/1475-7516/2012/04/034}, \href
  {http://adsabs.harvard.edu/abs/2012JCAP...04..034H} {4, 034},  \mn@eprint
  {arXiv} {1111.0052}

\bibitem[\protect\citeauthoryear{{Heitmann}, {Lawrence}, {Kwan}, {Habib}  \&
  {Higdon}}{{Heitmann} et~al.}{2014}]{2014ApJ...780..111H}
{Heitmann} K.,  {Lawrence} E.,  {Kwan} J.,  {Habib} S.,   {Higdon} D.,  2014,
  \mn@doi [\apj] {10.1088/0004-637X/780/1/111}, \href
  {http://ads.nao.ac.jp/abs/2014ApJ...780..111H} {780, 111},  \mn@eprint
  {arXiv} {1304.7849}

\bibitem[\protect\citeauthoryear{{Heymans} et~al.,}{{Heymans}
  et~al.}{2013}]{2013MNRAS.432.2433H}
{Heymans} C.,  et~al., 2013, \mn@doi [\mnras] {10.1093/mnras/stt601}, \href
  {http://adsabs.harvard.edu/abs/2013MNRAS.432.2433H} {432, 2433},  \mn@eprint
  {arXiv} {1303.1808}

\bibitem[\protect\citeauthoryear{{Hojjati}, {McCarthy}, {Harnois-Deraps}, {Ma},
  {Van Waerbeke}, {Hinshaw}  \& {Le Brun}}{{Hojjati}
  et~al.}{2015}]{2015JCAP...10..047H}
{Hojjati} A.,  {McCarthy} I.~G.,  {Harnois-Deraps} J.,  {Ma} Y.-Z.,  {Van
  Waerbeke} L.,  {Hinshaw} G.,   {Le Brun} A.~M.~C.,  2015, \mn@doi [\jcap]
  {10.1088/1475-7516/2015/10/047}, \href
  {http://adsabs.harvard.edu/abs/2015JCAP...10..047H} {10, 047},  \mn@eprint
  {arXiv} {1412.6051}

\bibitem[\protect\citeauthoryear{{Huff}, {Eifler}, {Hirata}, {Mandelbaum},
  {Schlegel}  \& {Seljak}}{{Huff} et~al.}{2014}]{2014MNRAS.440.1322H}
{Huff} E.~M.,  {Eifler} T.,  {Hirata} C.~M.,  {Mandelbaum} R.,  {Schlegel} D.,
   {Seljak} U.,  2014, \mn@doi [\mnras] {10.1093/mnras/stu145}, \href
  {http://adsabs.harvard.edu/abs/2014MNRAS.440.1322H} {440, 1322}

\bibitem[\protect\citeauthoryear{{Huterer} \& {Starkman}}{{Huterer} \&
  {Starkman}}{2003}]{2003PhRvL..90c1301H}
{Huterer} D.,  {Starkman} G.,  2003, \mn@doi [Physical Review Letters]
  {10.1103/PhysRevLett.90.031301}, \href
  {http://adsabs.harvard.edu/abs/2003PhRvL..90c1301H} {90, 031301},  \mn@eprint
  {} {astro-ph/0207517}

\bibitem[\protect\citeauthoryear{{Huterer} \& {Takada}}{{Huterer} \&
  {Takada}}{2005}]{2005APh....23..369H}
{Huterer} D.,  {Takada} M.,  2005, \mn@doi [Astroparticle Physics]
  {10.1016/j.astropartphys.2005.02.006}, \href
  {http://adsabs.harvard.edu/abs/2005APh....23..369H} {23, 369},  \mn@eprint {}
  {astro-ph/0412142}

\bibitem[\protect\citeauthoryear{{Huterer} \& {White}}{{Huterer} \&
  {White}}{2005}]{2005PhRvD..72d3002H}
{Huterer} D.,  {White} M.,  2005, \mn@doi [\prd] {10.1103/PhysRevD.72.043002},
  \href {http://adsabs.harvard.edu/abs/2005PhRvD..72d3002H} {72, 043002},
  \mn@eprint {} {astro-ph/0501451}

\bibitem[\protect\citeauthoryear{{Huterer}, {Takada}, {Bernstein}  \&
  {Jain}}{{Huterer} et~al.}{2006}]{2006MNRAS.366..101H}
{Huterer} D.,  {Takada} M.,  {Bernstein} G.,   {Jain} B.,  2006, \mn@doi
  [\mnras] {10.1111/j.1365-2966.2005.09782.x}, \href
  {http://adsabs.harvard.edu/abs/2006MNRAS.366..101H} {366, 101},  \mn@eprint
  {} {astro-ph/0506030}

\bibitem[\protect\citeauthoryear{{Jarvis} et~al.,}{{Jarvis}
  et~al.}{2015}]{2015arXiv150705603J}
{Jarvis} M.,  et~al., 2015, \href
  {http://adsabs.harvard.edu/abs/2015arXiv150705603J} {} \mn@eprint {arXiv}
  {1507.05603}

\bibitem[\protect\citeauthoryear{{Jing}, {Zhang}, {Lin}, {Gao}  \&
  {Springel}}{{Jing} et~al.}{2006}]{2006ApJ...640L.119J}
{Jing} Y.~P.,  {Zhang} P.,  {Lin} W.~P.,  {Gao} L.,   {Springel} V.,  2006,
  \mn@doi [\apjl] {10.1086/503547}, \href
  {http://adsabs.harvard.edu/abs/2006ApJ...640L.119J} {640, L119},  \mn@eprint
  {} {astro-ph/0512426}

\bibitem[\protect\citeauthoryear{{Joachimi}, {Schneider}  \&
  {Eifler}}{{Joachimi} et~al.}{2008}]{2008A&A...477...43J}
{Joachimi} B.,  {Schneider} P.,   {Eifler} T.,  2008, \mn@doi [\aap]
  {10.1051/0004-6361:20078400}, \href
  {http://ads.nao.ac.jp/abs/2008A%26A...477...43J} {477, 43},  \mn@eprint
  {arXiv} {0708.0387}

\bibitem[\protect\citeauthoryear{{Kadota}, {Dodelson}, {Hu}  \&
  {Stewart}}{{Kadota} et~al.}{2005}]{2005PhRvD..72b3510K}
{Kadota} K.,  {Dodelson} S.,  {Hu} W.,   {Stewart} E.~D.,  2005, \mn@doi [\prd]
  {10.1103/PhysRevD.72.023510}, \href
  {http://adsabs.harvard.edu/abs/2005PhRvD..72b3510K} {72, 023510},  \mn@eprint
  {} {astro-ph/0505158}

\bibitem[\protect\citeauthoryear{{Kaiser}, {Wilson}  \& {Luppino}}{{Kaiser}
  et~al.}{2000}]{2000astro.ph..3338K}
{Kaiser} N.,  {Wilson} G.,   {Luppino} G.~A.,  2000, \href
  {http://adsabs.harvard.edu/abs/2000astro.ph..3338K} {} \mn@eprint {}
  {astro-ph/0003338}

\bibitem[\protect\citeauthoryear{{Kitching} \& {Taylor}}{{Kitching} \&
  {Taylor}}{2011}]{2011MNRAS.416.1717K}
{Kitching} T.~D.,  {Taylor} A.~N.,  2011, \mn@doi [\mnras]
  {10.1111/j.1365-2966.2011.18772.x}, \href
  {http://adsabs.harvard.edu/abs/2011MNRAS.416.1717K} {416, 1717},  \mn@eprint
  {arXiv} {1012.3479}

\bibitem[\protect\citeauthoryear{{Kitching}, {Verde}, {Heavens}  \&
  {Jimenez}}{{Kitching} et~al.}{2016}]{2016MNRAS.459..971K}
{Kitching} T.~D.,  {Verde} L.,  {Heavens} A.~F.,   {Jimenez} R.,  2016, \mn@doi
  [\mnras] {10.1093/mnras/stw707}, \href
  {http://adsabs.harvard.edu/abs/2016MNRAS.459..971K} {459, 971},  \mn@eprint
  {arXiv} {1602.02960}

\bibitem[\protect\citeauthoryear{{Krause} \& {Eifler}}{{Krause} \&
  {Eifler}}{2016}]{2016arXiv160105779K}
{Krause} E.,  {Eifler} T.,  2016, \href
  {http://adsabs.harvard.edu/abs/2016arXiv160105779K} {} \mn@eprint {arXiv}
  {1601.05779}

\bibitem[\protect\citeauthoryear{{Krause} \& {Hirata}}{{Krause} \&
  {Hirata}}{2010}]{2010A&A...523A..28K}
{Krause} E.,  {Hirata} C.~M.,  2010, \mn@doi [\aap]
  {10.1051/0004-6361/200913524}, \href
  {http://ads.nao.ac.jp/abs/2010A%26A...523A..28K} {523, A28},  \mn@eprint
  {arXiv} {0910.3786}

\bibitem[\protect\citeauthoryear{{Krause}, {Eifler}  \& {Blazek}}{{Krause}
  et~al.}{2016}]{2016MNRAS.456..207K}
{Krause} E.,  {Eifler} T.,   {Blazek} J.,  2016, \mn@doi [\mnras]
  {10.1093/mnras/stv2615}, \href
  {http://adsabs.harvard.edu/abs/2016MNRAS.456..207K} {456, 207},  \mn@eprint
  {arXiv} {1506.08730}

\bibitem[\protect\citeauthoryear{{Kuijken} et~al.,}{{Kuijken}
  et~al.}{2015}]{2015MNRAS.454.3500K}
{Kuijken} K.,  et~al., 2015, \mn@doi [\mnras] {10.1093/mnras/stv2140}, \href
  {http://adsabs.harvard.edu/abs/2015MNRAS.454.3500K} {454, 3500},  \mn@eprint
  {arXiv} {1507.00738}

\bibitem[\protect\citeauthoryear{{Laureijs} et~al.,}{{Laureijs}
  et~al.}{2011}]{2011arXiv1110.3193L}
{Laureijs} R.,  et~al., 2011, \href
  {http://adsabs.harvard.edu/abs/2011arXiv1110.3193L} {} \mn@eprint {arXiv}
  {1110.3193}

\bibitem[\protect\citeauthoryear{{Lewandowski}, {Perko}  \&
  {Senatore}}{{Lewandowski} et~al.}{2015}]{2015JCAP...05..019L}
{Lewandowski} M.,  {Perko} A.,   {Senatore} L.,  2015, \mn@doi [\jcap]
  {10.1088/1475-7516/2015/05/019}, \href
  {http://adsabs.harvard.edu/abs/2015JCAP...05..019L} {5, 019},  \mn@eprint
  {arXiv} {1412.5049}

\bibitem[\protect\citeauthoryear{{Lewis}, {Challinor}  \& {Lasenby}}{{Lewis}
  et~al.}{2000}]{camb}
{Lewis} A.,  {Challinor} A.,   {Lasenby} A.,  2000, \mn@doi [\apj]
  {10.1086/309179}, \href {http://adsabs.harvard.edu/abs/2000ApJ...538..473L}
  {538, 473},  \mn@eprint {} {astro-ph/9911177}

\bibitem[\protect\citeauthoryear{{Lin} et~al.,}{{Lin}
  et~al.}{2012}]{2012ApJ...761...15L}
{Lin} H.,  et~al., 2012, \mn@doi [\apj] {10.1088/0004-637X/761/1/15}, \href
  {http://adsabs.harvard.edu/abs/2012ApJ...761...15L} {761, 15},  \mn@eprint
  {arXiv} {1111.6622}

\bibitem[\protect\citeauthoryear{{Ma}, {Van Waerbeke}, {Hinshaw}, {Hojjati},
  {Scott}  \& {Zuntz}}{{Ma} et~al.}{2015}]{2015JCAP...09..046M}
{Ma} Y.-Z.,  {Van Waerbeke} L.,  {Hinshaw} G.,  {Hojjati} A.,  {Scott} D.,
  {Zuntz} J.,  2015, \mn@doi [\jcap] {10.1088/1475-7516/2015/09/046}, \href
  {http://adsabs.harvard.edu/abs/2015JCAP...09..046M} {9, 046},  \mn@eprint
  {arXiv} {1404.4808}

\bibitem[\protect\citeauthoryear{{MacCrann}}{{MacCrann}}{2016}]{maccrann-thesis}
{MacCrann} N.,  2016, PhD thesis, University of Manchester

\bibitem[\protect\citeauthoryear{{MacCrann}, {Zuntz}, {Bridle}, {Jain}  \&
  {Becker}}{{MacCrann} et~al.}{2015}]{2015MNRAS.451.2877M}
{MacCrann} N.,  {Zuntz} J.,  {Bridle} S.,  {Jain} B.,   {Becker} M.~R.,  2015,
  \mn@doi [\mnras] {10.1093/mnras/stv1154}, \href
  {http://adsabs.harvard.edu/abs/2015MNRAS.451.2877M} {451, 2877},  \mn@eprint
  {arXiv} {1408.4742}

\bibitem[\protect\citeauthoryear{{MacCrann} et~al.}{{MacCrann}
  et~al.}{2016}]{maccrann-inprep}
{MacCrann} N.,  et~al., 2016, in preparation

\bibitem[\protect\citeauthoryear{{Mead}, {Peacock}, {Heymans}, {Joudaki}  \&
  {Heavens}}{{Mead} et~al.}{2015}]{2015MNRAS.454.1958M}
{Mead} A.~J.,  {Peacock} J.~A.,  {Heymans} C.,  {Joudaki} S.,   {Heavens}
  A.~F.,  2015, \mn@doi [\mnras] {10.1093/mnras/stv2036}, \href
  {http://adsabs.harvard.edu/abs/2015MNRAS.454.1958M} {454, 1958},  \mn@eprint
  {arXiv} {1505.07833}

\bibitem[\protect\citeauthoryear{{Mohammed} \& {Seljak}}{{Mohammed} \&
  {Seljak}}{2014}]{2014MNRAS.445.3382M}
{Mohammed} I.,  {Seljak} U.,  2014, \mn@doi [\mnras] {10.1093/mnras/stu1972},
  \href {http://adsabs.harvard.edu/abs/2014MNRAS.445.3382M} {445, 3382},
  \mn@eprint {arXiv} {1407.0060}

\bibitem[\protect\citeauthoryear{{Mohammed}, {Martizzi}, {Teyssier}  \&
  {Amara}}{{Mohammed} et~al.}{2014}]{2014arXiv1410.6826M}
{Mohammed} I.,  {Martizzi} D.,  {Teyssier} R.,   {Amara} A.,  2014, \href
  {http://adsabs.harvard.edu/abs/2014arXiv1410.6826M} {} \mn@eprint {arXiv}
  {1410.6826}

\bibitem[\protect\citeauthoryear{{Natarajan}, {Zentner}, {Battaglia}  \&
  {Trac}}{{Natarajan} et~al.}{2014}]{2014PhRvD..90f3516N}
{Natarajan} A.,  {Zentner} A.~R.,  {Battaglia} N.,   {Trac} H.,  2014, \mn@doi
  [\prd] {10.1103/PhysRevD.90.063516}, \href
  {http://adsabs.harvard.edu/abs/2014PhRvD..90f3516N} {90, 063516},  \mn@eprint
  {arXiv} {1405.6205}

\bibitem[\protect\citeauthoryear{{Oguri} \& {Takada}}{{Oguri} \&
  {Takada}}{2011}]{2011PhRvD..83b3008O}
{Oguri} M.,  {Takada} M.,  2011, \mn@doi [\prd] {10.1103/PhysRevD.83.023008},
  \href {http://adsabs.harvard.edu/abs/2011PhRvD..83b3008O} {83, 023008},
  \mn@eprint {arXiv} {1010.0744}

\bibitem[\protect\citeauthoryear{{Osato}, {Shirasaki}  \& {Yoshida}}{{Osato}
  et~al.}{2015}]{2015ApJ...806..186O}
{Osato} K.,  {Shirasaki} M.,   {Yoshida} N.,  2015, \mn@doi [\apj]
  {10.1088/0004-637X/806/2/186}, \href
  {http://adsabs.harvard.edu/abs/2015ApJ...806..186O} {806, 186},  \mn@eprint
  {arXiv} {1501.02055}

\bibitem[\protect\citeauthoryear{{Planck Collaboration} et~al.,}{{Planck
  Collaboration} et~al.}{2015}]{2015arXiv150201589P}
{Planck Collaboration} et~al., 2015, \href
  {http://adsabs.harvard.edu/abs/2015arXiv150201589P} {} \mn@eprint {arXiv}
  {1502.01589}

\bibitem[\protect\citeauthoryear{{Rudd}, {Zentner}  \& {Kravtsov}}{{Rudd}
  et~al.}{2008}]{2008ApJ...672...19R}
{Rudd} D.~H.,  {Zentner} A.~R.,   {Kravtsov} A.~V.,  2008, \mn@doi [\apj]
  {10.1086/523836}, \href {http://adsabs.harvard.edu/abs/2008ApJ...672...19R}
  {672, 19},  \mn@eprint {} {astro-ph/0703741}

\bibitem[\protect\citeauthoryear{{Samsing}, {Linder}  \& {Smith}}{{Samsing}
  et~al.}{2012}]{2012PhRvD..86l3504S}
{Samsing} J.,  {Linder} E.~V.,   {Smith} T.~L.,  2012, \mn@doi [\prd]
  {10.1103/PhysRevD.86.123504}, \href
  {http://adsabs.harvard.edu/abs/2012PhRvD..86l3504S} {86, 123504},  \mn@eprint
  {arXiv} {1208.4845}

\bibitem[\protect\citeauthoryear{{Schaye} et~al.,}{{Schaye}
  et~al.}{2010}]{2010MNRAS.402.1536S}
{Schaye} J.,  et~al., 2010, \mn@doi [\mnras]
  {10.1111/j.1365-2966.2009.16029.x}, \href
  {http://adsabs.harvard.edu/abs/2010MNRAS.402.1536S} {402, 1536},  \mn@eprint
  {arXiv} {0909.5196}

\bibitem[\protect\citeauthoryear{{Schmidt}, {Rozo}, {Dodelson}, {Hui}  \&
  {Sheldon}}{{Schmidt} et~al.}{2009}]{2009ApJ...702..593S}
{Schmidt} F.,  {Rozo} E.,  {Dodelson} S.,  {Hui} L.,   {Sheldon} E.,  2009,
  \mn@doi [\apj] {10.1088/0004-637X/702/1/593}, \href
  {http://ads.nao.ac.jp/abs/2009ApJ...702..593S} {702, 593},  \mn@eprint
  {arXiv} {0904.4703}

\bibitem[\protect\citeauthoryear{{Schneider} \& {Teyssier}}{{Schneider} \&
  {Teyssier}}{2015}]{2015JCAP...12..049S}
{Schneider} A.,  {Teyssier} R.,  2015, \mn@doi [\jcap]
  {10.1088/1475-7516/2015/12/049}, \href
  {http://adsabs.harvard.edu/abs/2015JCAP...12..049S} {12, 049},  \mn@eprint
  {arXiv} {1510.06034}

\bibitem[\protect\citeauthoryear{{Semboloni}, {Hoekstra}, {Schaye}, {van
  Daalen}  \& {McCarthy}}{{Semboloni} et~al.}{2011}]{2011MNRAS.417.2020S}
{Semboloni} E.,  {Hoekstra} H.,  {Schaye} J.,  {van Daalen} M.~P.,   {McCarthy}
  I.~G.,  2011, \mn@doi [\mnras] {10.1111/j.1365-2966.2011.19385.x}, \href
  {http://ads.nao.ac.jp/abs/2011MNRAS.417.2020S} {417, 2020},  \mn@eprint
  {arXiv} {1105.1075}

\bibitem[\protect\citeauthoryear{{Semboloni}, {Hoekstra}  \&
  {Schaye}}{{Semboloni} et~al.}{2013}]{2013MNRAS.434..148S}
{Semboloni} E.,  {Hoekstra} H.,   {Schaye} J.,  2013, \mn@doi [\mnras]
  {10.1093/mnras/stt1013}, \href {http://ads.nao.ac.jp/abs/2013MNRAS.434..148S}
  {434, 148},  \mn@eprint {arXiv} {1210.7303}

\bibitem[\protect\citeauthoryear{{Shapiro}}{{Shapiro}}{2009}]{2009ApJ...696..775S}
{Shapiro} C.,  2009, \mn@doi [\apj] {10.1088/0004-637X/696/1/775}, \href
  {http://ads.nao.ac.jp/abs/2009ApJ...696..775S} {696, 775},  \mn@eprint
  {arXiv} {0812.0769}

\bibitem[\protect\citeauthoryear{{Simon}}{{Simon}}{2012}]{2012A&A...543A...2S}
{Simon} P.,  2012, \mn@doi [\aap] {10.1051/0004-6361/201118224}, \href
  {http://adsabs.harvard.edu/abs/2012A%26A...543A...2S} {543, A2},  \mn@eprint
  {arXiv} {1202.2046}

\bibitem[\protect\citeauthoryear{{Smith} et~al.,}{{Smith}
  et~al.}{2003}]{halofit}
{Smith} R.~E.,  et~al., 2003, \mn@doi [\mnras]
  {10.1046/j.1365-8711.2003.06503.x}, \href
  {http://adsabs.harvard.edu/abs/2003MNRAS.341.1311S} {341, 1311},  \mn@eprint
  {} {astro-ph/0207664}

\bibitem[\protect\citeauthoryear{{Takahashi}, {Sato}, {Nishimichi}, {Taruya}
  \& {Oguri}}{{Takahashi} et~al.}{2012}]{halofit-update}
{Takahashi} R.,  {Sato} M.,  {Nishimichi} T.,  {Taruya} A.,   {Oguri} M.,
  2012, \mn@doi [\apj] {10.1088/0004-637X/761/2/152}, \href
  {http://adsabs.harvard.edu/abs/2012ApJ...761..152T} {761, 152},  \mn@eprint
  {arXiv} {1208.2701}

\bibitem[\protect\citeauthoryear{{The Dark Energy Survey Collaboration}
  et~al.,}{{The Dark Energy Survey Collaboration}
  et~al.}{2015}]{2015arXiv150705552T}
{The Dark Energy Survey Collaboration} et~al., 2015, \href
  {http://adsabs.harvard.edu/abs/2015arXiv150705552T} {} \mn@eprint {arXiv}
  {1507.05552}

\bibitem[\protect\citeauthoryear{{Van Daalen}, {Schaye}, {Booth}  \& {Dalla
  Vecchia}}{{Van Daalen} et~al.}{2011}]{2011MNRAS.415.3649V}
{Van Daalen} M.~P.,  {Schaye} J.,  {Booth} C.~M.,   {Dalla Vecchia} C.,  2011,
  \mn@doi [\mnras] {10.1111/j.1365-2966.2011.18981.x}, \href
  {http://adsabs.harvard.edu/abs/2011MNRAS.415.3649V} {415, 3649},  \mn@eprint
  {arXiv} {1104.1174}

\bibitem[\protect\citeauthoryear{{Van Waerbeke} et~al.,}{{Van Waerbeke}
  et~al.}{2000}]{2000A&A...358...30V}
{Van Waerbeke} L.,  et~al., 2000, \aap, \href
  {http://ads.nao.ac.jp/abs/2000A%26A...358...30V} {358, 30},  \mn@eprint {}
  {astro-ph/0002500}

\bibitem[\protect\citeauthoryear{{Van Waerbeke}, {Hinshaw}  \& {Murray}}{{Van
  Waerbeke} et~al.}{2014}]{2014PhRvD..89b3508V}
{Van Waerbeke} L.,  {Hinshaw} G.,   {Murray} N.,  2014, \mn@doi [\prd]
  {10.1103/PhysRevD.89.023508}, \href
  {http://adsabs.harvard.edu/abs/2014PhRvD..89b3508V} {89, 023508},  \mn@eprint
  {arXiv} {1310.5721}

\bibitem[\protect\citeauthoryear{{Viola} et~al.,}{{Viola}
  et~al.}{2015}]{2015MNRAS.452.3529V}
{Viola} M.,  et~al., 2015, \mn@doi [\mnras] {10.1093/mnras/stv1447}, \href
  {http://adsabs.harvard.edu/abs/2015MNRAS.452.3529V} {452, 3529},  \mn@eprint
  {arXiv} {1507.00735}

\bibitem[\protect\citeauthoryear{{Vogelsberger} et~al.,}{{Vogelsberger}
  et~al.}{2014}]{2014Natur.509..177V}
{Vogelsberger} M.,  et~al., 2014, \mn@doi [\nat] {10.1038/nature13316}, \href
  {http://adsabs.harvard.edu/abs/2014Natur.509..177V} {509, 177},  \mn@eprint
  {arXiv} {1405.1418}

\bibitem[\protect\citeauthoryear{{White}}{{White}}{2004}]{2004APh....22..211W}
{White} M.,  2004, \mn@doi [Astroparticle Physics]
  {10.1016/j.astropartphys.2004.06.001}, \href
  {http://adsabs.harvard.edu/abs/2004APh....22..211W} {22, 211},  \mn@eprint {}
  {astro-ph/0405593}

\bibitem[\protect\citeauthoryear{{Wittman}, {Tyson}, {Kirkman}, {Dell'Antonio}
  \& {Bernstein}}{{Wittman} et~al.}{2000}]{2000Natur.405..143W}
{Wittman} D.~M.,  {Tyson} J.~A.,  {Kirkman} D.,  {Dell'Antonio} I.,
  {Bernstein} G.,  2000, \nat, \href
  {http://ads.nao.ac.jp/abs/2000Natur.405..143W} {405, 143},  \mn@eprint {}
  {astro-ph/0003014}

\bibitem[\protect\citeauthoryear{{Zentner}, {Rudd}  \& {Hu}}{{Zentner}
  et~al.}{2008}]{2008PhRvD..77d3507Z}
{Zentner} A.~R.,  {Rudd} D.~H.,   {Hu} W.,  2008, \mn@doi [\prd]
  {10.1103/PhysRevD.77.043507}, \href
  {http://ads.nao.ac.jp/abs/2008PhRvD..77d3507Z} {77, 043507},  \mn@eprint
  {arXiv} {0709.4029}

\bibitem[\protect\citeauthoryear{{Zentner}, {Semboloni}, {Dodelson}, {Eifler},
  {Krause}  \& {Hearin}}{{Zentner} et~al.}{2013}]{2013PhRvD..87d3509Z}
{Zentner} A.~R.,  {Semboloni} E.,  {Dodelson} S.,  {Eifler} T.,  {Krause} E.,
  {Hearin} A.~P.,  2013, \mn@doi [\prd] {10.1103/PhysRevD.87.043509}, \href
  {http://ads.nao.ac.jp/abs/2013PhRvD..87d3509Z} {87, 043509},  \mn@eprint
  {arXiv} {1212.1177}

\bibitem[\protect\citeauthoryear{{Zhan} \& {Knox}}{{Zhan} \&
  {Knox}}{2004}]{2004ApJ...616L..75Z}
{Zhan} H.,  {Knox} L.,  2004, \mn@doi [\apjl] {10.1086/426712}, \href
  {http://adsabs.harvard.edu/abs/2004ApJ...616L..75Z} {616, L75},  \mn@eprint
  {} {astro-ph/0409198}

\bibitem[\protect\citeauthoryear{{Zuntz} et~al.,}{{Zuntz}
  et~al.}{2015}]{2015A&C....12...45Z}
{Zuntz} J.,  et~al., 2015, \mn@doi [Astronomy and Computing]
  {10.1016/j.ascom.2015.05.005}, \href
  {http://adsabs.harvard.edu/abs/2015A%26C....12...45Z} {12, 45},  \mn@eprint
  {arXiv} {1409.3409}

\makeatother
\end{thebibliography}

\bsp	
\label{lastpage}
\end{document}